\documentclass[twoside,11pt]{article}

\usepackage{blindtext}
\usepackage{tabularx}
\usepackage{amsmath}
\usepackage{xcolor}
\usepackage{booktabs}
\usepackage{subcaption}
\usepackage{graphicx}

\usepackage{irrj2e}

\newcommand{\dataset}{{\cal D}}
\newcommand{\fracpartial}[2]{\frac{\partial #1}{\partial  #2}}

\newcommand{\new}[1]{\textcolor{black}{#1}}
\newcommand{\newreview}[1]{\textcolor{black}{#1}}

\newcommand{\shrink}{\vspace*{-.9\baselineskip}}

\usepackage{lastpage}

\irrjheading{1}{2025}{1-1}{20-05-2025}{date published}{paper id}{Emma J. Gerritse, Faegheh Hasibi, Arjen P. de Vries}
\ShortHeadings{Graph-Embedding Empowered Entity Retrieval}{Gerritse, Hasibi, and de Vries}
\firstpageno{1}

\begin{document}

\title{Graph Embeddings to Empower Entity Retrieval}

\author{\name Emma J. Gerritse \email emma.gerritse@ru.nl \\
       \addr 
       % Institute for Comutign and Information Sciences\\
       Radboud University\\
       The Netherlands
       \AND
       \name Faegheh Hasibi \email faegheh.hasibi@ru.nl \\
       \addr 
       % Institute for Comutign and Information Sciences\\
       Radboud University\\
       The Netherlands
       \AND
       \name Arjen P. de Vries \email arjen.devries@ru.nl \\
       \addr 
       % Institute for Comutign and Information Sciences\\
       Radboud University\\
       The Netherlands}

\editor{My editor}

\maketitle

\begin{abstract}
In this research, we investigate methods for entity retrieval using graph embeddings. While various methods have been proposed over the years, most utilize a single graph embedding and entity linking approach. This hinders our understanding of how different graph embedding and entity linking methods impact entity retrieval.
%methods
To address this gap, we investigate the effects of three different categories of graph embedding techniques and five different entity linking methods.  We perform a reranking of entities using the distance between the embeddings of annotated entities and the entities we wish to rerank. 
%results
% We find that Wikipedia2Vec graph embeddings, which utilize both graph structure and textual descriptions of entities, are the most effective for entity retireval. \todo{The entity annotations provided by SMAPH, TagMe and ELQ lead to the best scores in this reranking method. We also introduce a new set of human-curated annotations for DBpedia-entity V2, focusing on annotating concepts and named entities. We find that the human-curated entity annotations focussing on concepts perform better on DBpedia-entity V2 than the annotations focussing on named entities. }
% conclusion
% We conclude that choices in both entity linkers and graph embeddings matter when using graph embeddings and entity linkers for entity retrieval. 
We conclude that the selection of both graph embeddings and entity linkers significantly impacts the effectiveness of entity retrieval.
For graph embeddings,  methods that incorporate both graph structure and textual descriptions of entities are the most effective. 
For entity linking, both precision and recall concerning concepts are important for optimal retrieval performance.  Additionally, it is essential for the graph to encompass as many entities as possible.

\end{abstract}


\section{Introduction}
\label{sec:intro}

% entity/word embeddings are popular 
% we wonder if entity embeddings help entity retrieval
% do embeddings help or does the graph structure help?
% we show that graph embeddings help
% we show a new use for dbpedia entity collection

%\textbf{Importance of KG \& Entity for retrieval task}

 %informarion needs are often entity oriented
 %either the query contains entities, or the answer should be an entity
 %

A significant portion of user queries in web search and question answering explicitly or implicitly reference entities, with users either asking for some entity-related facts or looking for the page of a specific entity \citep{balog2018entity, meij:2014:entity}.
%Users may benefit from entity information to satisfy their information need when looking for information. More and more, queries feature entities, or users are looking for the page of a specific entity \citep{meij:2014:entity}. 
Entities are a vital part of information retrieval research, and this has led to the development of datasets  \citep{Hasibi:2017:DVT} and methods for effective entity retrieval \citep{ Arabzadeh:2024:LaQuE, Gerritse:2022:Entity, chatterjee:2022:berter, Gerritse:2020:GEEER, dietz:2019:ENT, Garigliotti2019}. %Jafarzadeh:2024:LCR, , , Dalton:2014:EQF %\todo{cite a bunch of older ER papers}
% Zamiri:2024:BNA,
%\textbf{What are KG embeddings, what do they capture, why are they important}
Entities are commonly stored in a Knowledge Graph (KG), which connects them through various relationships. The entire knowledge graph can then be represented using graph embeddings, which capture the rich and human-curated information in low-dimensional vector spaces, representing the similarity and relations between different entities. Popularized by methods such as Trans-E \citep{bordes2013translating} and Wikipedia2Vec \citep{Yamada:2018:WOT}, graph embeddings have proven to be useful for tasks like link prediction \citep{bordes2013translating} and text classification \citep{yamada:2019:neural}.

%\textbf{The use of KG embedding in entity Retrieval}

\newreview{This study investigates the use of KG embeddings to enhance lexical entity retrieval models.
While recent transformer-based entity retrieval models have demonstrated superior performance compared to KG-augmented lexical models~\citep{chatterjee:2022:berter, Gerritse:2022:Entity}, they still struggle with long-tail entities~\citep{Gerritse:2022:Entity}. Similarly, dense retrievers face challenges in generalizing to the entity retrieval task~\citep{Kamalloo:2024:RepBEIR}, performing worse than or only on par with strong lexical retrieval models such as BM25F or GEEER~\citep{Gerritse:2020:GEEER}.
These findings are in line with recent studies that show LLMs struggle to answer factual questions about long tail entities, and Retrieval Augmented Generation (RAG) needs to be employed to fill this gap~\citep{mallen:2023:trust,Soudani:2024:FTvsRAG}. }

%\textbf{re-ranking using static word embeddings vs entity embeddings}
Various methods have been proposed to include graph embeddings for entity retrieval. \citet{Gerritse:2020:GEEER} and \citet{Nikolaev:2020:joint} introduce methods using the similarity between the graph embedding vectors to compute relevance between queries and entities. These fairly simple methods improve over the previously established state of the art.
%\textbf{re-ranking using transformer models}
% With the advancements of transformer models and pre-trained language models, recent works opt to enhance transformer architectures with graph embeddings \citep{daza:2021:inductive,Gerritse:2022:Entity}, or combine the previously discussed similarity-based methods with transformer architectures \citep{Oza:2023:Entity}. 
However, existing approaches for utilizing graph embeddings for entity retrieval typically employ a single entity linker and embedding methods. Therefore, the effect of different entity linking and graph embedding methods on overall entity retrieval performance has remained unexplored.
%to our knowledge, no research has been conducted to compare the effect of different entity linking and graph embedding methods on overall retrieval performance. 

In this study, we aim to fill this gap and explore various aspects of utilizing knowledge graph embeddings for entity retrieval. We base our approach on the method proposed by \citet{Gerritse:2020:GEEER}, an early work on the use of graph embeddings for entity retrieval that has been reproduced and extended in various setups \citep{Oza:2023:Entity, chatterjee:2022:berter, Jafarzadeh:2022:LRK, daza:2021:inductive}. Additionally, it provides a simple framework for the fusion of graph embeddings and retrieval models, enabling us to understand the effect of different entity linking and embedding methods on the entity retrieval task.

%Even though transformer-based methods tend to outperform most older methods, it is still necessary to understand the more straightforward methods that deploy entities and graph embeddings for several reasons. First, these graph-embedding methods have a significantly lower runtime in training and inference than transformer-based methods. Second, the best-performing transformer-based methods for entity retrieval, as discussed above, still rely heavily on entity linkers and graph embeddings. However, to our knowledge, most of these papers utilize a single entity linker and graph embedding method, and no research has been done comparing the differences between entity linkers and graph embedding methods. 

%In this study, we explore various questions in utilizing kg embeddings for ER and We follow cit an we use a kg embedding re-ranking method to explore these RQs

We tackle the following research questions in this work:

% \textbf{RQ 1.} \textit{Which entity linking methods are most suitable for information retrieval using entity embeddings?} 
% Different entity-linking methods with various goals have been introduced throughout the years. For instance, REL \citep{vanHulst:2020:rel} provides high precision of linked entities, while methods like SMAPH \citep{Cornolti:2018:SMAPH} and SMPAH leverage pre-existing methods. In this work, we seek to find how the differences in entity linking methods and the produced annotations affect the entity retrieval performance. \todo{We find that the entity-linking methods that work best for entity retrieval are SMAPH, TagMe, and ELQ.} 

% \textbf{RQ 1.} \textit{Which properties are most important for entity linking methods for information retrieval?}
\noindent
\textbf{RQ 1.} \textit{
\newreview{How do different entity linking methods and their specific properties affect graph embedding–empowered entity retrieval?}}
Different entity-linking methods have been introduced throughout the years. Some take into account all different scenarios that might be meant by a query, for example, considering all other meanings of homonyms and polysemes, while others consider the most probable interpretation of a query and map each entity mention to a single entity.   Additionally, entity linking methods such as REL \citep{vanHulst:2020:rel} and Nordlys \citep{Hasibi:2017:NTE} provide high precision of named entities like people, location, and organizations. In contrast,  methods like SMAPH \citep{Cornolti:2018:SMAPH} and TagMe \citep{ferragina2010tagme} are designed to annotate both named entities and general concepts, such as \textsc{Democracy}.
We hypothesize that when using graph embeddings for entity retrieval, it is beneficial to include as many entity embeddings relevant to the query as possible, encompassing both concepts and name entities. To assess this hypothesis, we manually annotate queries of the DBpedia-Entity collection \citep{Hasibi:2017:DVT}, and compare different entity linkers using this ground truth. We find that, indeed, the best entity retrieval performance is achieved with entity linkers that annotate both concepts and name entities.
% best-performing entity linkers are the ones that also include concepts.  thus also having to annotate concepts.
% Following the previous research question, we aim to investigate what might cause variations in performance when using different entity-linking methods. For this, we compare two approaches to annotating queries: the first is taking into account all different scenarios that might be meant by a query, for example, considering all other meanings of homonyms and polysemes. The second is including concepts in the linked entities. Typically, entity linkers focus on linking named entities; concepts are often omitted (e.g. the entities `country' or `actor' will not be linked, but specific named entities like `The United States of America' or `Keanu Reeves' will be). However, we hypothesize that when using graph embeddings, it is better to have as many entity embeddings relevant to the query as possible, thus also having to annotate concepts. We find that, indeed, the best-performing entity linkers are the ones that also include concepts. 

\textbf{RQ 2.} \textit{\newreview{Which graph embedding techniques work best for entity retrieval methods empowered by graph embeddings?}}
In addition to diverse approaches proposed for entity linking,  numerous algorithms have been introduced for graph embeddings. This work classifies them into three distinct groups, namely skip-gram-based \citep{Yamada:2016:JLE}, transition-based \citep{Trouillon:2016:Complex}, and random-walk-based \citep{Ristoski:2016:rdf2vec}. Often, research in the field of information retrieval only utilizes one type of graph embedding without considering the other classes of methods. In this work, we compare these various methods. We find that Wikipedia2Vec, as a skip-gram-based method, has the highest performance of all methods, provided that much effort is made to correct missing entities. We also see a positive impact from including as many entities as possible in the graph embedding for entity retrieval,  even if it significantly increases the graph size during embedding training.

\textbf{RQ 3.} \newreview{\textit{How does the structural information captured by the skip-gram–based graph embedding approach, Wikipedia2Vec, contribute to entity retrieval effectiveness?}}
% We investigate how the structural information captured in graph embeddings can contribute to improved retrieval effectiveness in entity-oriented search. 
To address this research question, we train two versions of Wikipedia2Vec embeddings,  with and without link graph, and compare the obtained embeddings and retrieval results. Utilizing the cluster hypothesis \citep{rijsbergen1979information}, we show that a representation of the graph structure in the embeddings leads to better clusters and higher effectiveness of retrieval results. We further see that queries \newreview{with correctly linked entities (by an entity linker)} are helped the most, while queries with wrongly linked entities are helped the least.

This work is an extension of ~\citep{Gerritse:2020:GEEER}, which makes the following new contributions: (i) We provide a comprehensive investigation of different entity linking and graph embedding methods for entity retrieval, (ii) We provide a new set of entity annotations of the widely-used DBpedia-Entity collection that includes both concepts and named entities.
%, (iii)  We perform an extensive analysis of entity embeddings and investigate how to explain our result by comparing the structure of the two variations of Wikipedia2Vec embeddings. We also analyze why some queries are helped while others are not. 
All the resources developed in the course of this study are made publicly available at~\url{https://github.com/informagi/GEEER}.

\section{Related Work}
\label{sec:related}

\subsection{Word and Entity Embeddings}

Distributional representations of language have been the object of study for many years in natural language processing (NLP), because of their promise to represent words not in isolation, but semantically, with their immediate context. 
Algorithms like Word2Vec \citep{Mikolov:2013:DRW} and Glove \citep{pennington2014glove} construct a vector space of word domains where similar words are mapped together (based on their linguistic context).
Word2Vec embeddings are extracted from neural networks that predict words based on their context (continuous bag of words) or that predict the context for a given word (skip-gram).
%Perhaps due to the large amounts of data that can be processed nowadays, these word embedding representations have turned out to be highly effective in a wide variety of natural language processing tasks.
These word-embedding representations have proven to be highly effective in various Information Retrieval (IR) and NLP tasks.

Word embeddings have been shown to improve effectiveness in document retrieval \citep{dehghani2017neural,diaz2016expansion}. 
In \citep{diaz2016expansion}, locally trained word embeddings are used for query expansion. 
Here, queries are expanded with terms highly similar to the query, and it is shown that this method beats several other neural methods. In \citep{dehghani2017neural}, embeddings are used to train a neural ranking model using weak supervision. 
The authors use query embeddings and document embeddings to predict relevance between queries and documents when given BM25 scores as labels, outperforming BM25.  

% Word embeddings capture the immediate linguistic context of word occurrences. 
% Going beyond the text itself, researchers have proposed developing  \emph{graph embeddings} to encode not only words in text, but also their contextual relationships within semi-structured documents represented as graphs. For example, this approach can help distinguish the occurrence of a word in the title of a document from its occurrences in a paragraph, or in a document's anchor text.

%Different methods of producing graph embeddings have been proposed, which we categorize into three groups: transition-based, skip-gram-based, and random-walk-based approaches.

\newreview{
Word embeddings capture the immediate linguistic context of word occurrences. 
Going beyond the text itself, researchers across various research communities have proposed a vast number of Knowledge Graph (KG) embedding methods. A KG is a graph where the nodes represent entities, and the edges represent the relations between them.}
% \todo{We talk about three categories of graph embeddings (skip-gram, random wal, and transition-based) but they are not explained here. Also RDF2Vec is not explained. Please reorganize and add.}
One of the earliest and most well-known KG embedding methods is TransE \citep{bordes2013translating}, which falls under the transition-based category. The TransE method considers graph edges as  \textit{(head, label, tail)} triples, where \textit{label} is the value of the edge. 
Under the objective that adding graph embedding vectors of the \textit{head} and the \textit{label} should result in the vector of the \textit{tail}, these embeddings are learned by gradient descent. TransE has been influential but has proved to be ineffective for anti-symmetric relations present in the knowledge graph. This resulted in various proposals for KG embeddings that extend this method with additional objectives, such as DistMult~\citep{yang:2015:embedding}, ComplEx~\citep{Trouillon:2016:Complex}, and SimplE~\citep{Kazemi:2018:simple}. 

\newreview{ComplEx~\citep{Trouillon:2016:Complex} is a robust and widely used graph embedding, which utilizes complex vectors instead of real vectors and is better suited to represent anti-symmetric relations.  Several studies~\citep{Ruffinelli:2020:you, Chekalina:2022:meker, Kochsiek:2023:friendly} have demonstrated that ComplEx achieves competitive performance on the Wikidata5M dataset\citep{Wang:2021:kepler}.  In particular, \cite{Ruffinelli:2020:you} performed parameter tuning of ComplEx on the Wikidata5M dataset\footnote{highlighted in the GitHub repository \url{https://github.com/uma-pi1/kge}, last accessed 20-05-2025}, and showed that when training conditions are standardized, models such as TransE, ComplEx, and DistMult perform similarly. }

% \newreview{A more comparable benchmark to Wikipedia in the graph embedding field is the Wikidata5M~\citep{Wang:2021:kepler} dataset. Several studies~\citep{Ruffinelli:2020:you, Chekalina:2022:meker, Kochsiek:2023:friendly} have demonstrated that ComplEx achieves competitive performance on Wikidata5M. In particular, \citet{Ruffinelli:2020:you} performed parameter tuning of ComplEx on Wikidata5M\footnote{as highlighted in the official GitHub repository \url{https://github.com/uma-pi1/kge}} and showed that  when training conditions are standardized, models such as TransE, ComplEx, and DistMult perform similarly.
% Several studies~\citep{Ruffinelli:2020:you, Chekalina:2022:meker, Kochsiek:2023:friendly} have demonstrated that  In particular, \cite{Ruffinelli:2020:you} performed parameter tuning of ComplEx on the Wikidata5M dataset\footnote{as highlighted in the official GitHub repository \url{https://github.com/uma-pi1/kge}} \citep{Wang:2021:kepler}, and showed that when training conditions are standardized, models such as TransE, ComplEx, and DistMult perform similarly. \cite{Ruffinelli:2020:you} include both early (TransE) and advanced (ComplEx) translation-based KG embedding models.}

% \newreview{Most of these approaches are evaluated on link prediction tasks on FB15k-237 and WN18RR datasets. However, these datasets are considerably smaller than Wikipedia: FB15k contains only 1,345 relations and 14,951 entities, and WN18 includes just 18 relations and 40,943 entities. Given this substantial difference in scale and complexity, results obtained from these datasets may not generalize to Wikipedia.}

\newreview{Another class of KG embeddings that has been widely applied in downstream tasks~\citep{Oza:2023:Entity, Gerritse:2022:Entity}, without direct comparison with translation-based embeddings in the literature, is skip-gram-based approaches, such as Deepwalk~\citep{perozzi2014deepwalk} and RDF2Vec~\citep{Ristoski:2016:rdf2vec}.}
Deepwalk \citep{perozzi2014deepwalk} is a graph embedding method that expects non-labeled edges. It first randomly samples vertices from the graph as starting points, and then performs a random walk from each of these starting points. The vertices walked in these random walks can then be represented as sentences. After having created these sentences from the graph, Deepwalk uses these as input for a Word2Vec-based approach using the skip-gram method. RDF2Vec \citep{Ristoski:2016:rdf2vec} extends Deepwalk to work on knowledge graphs, by not only using the vertices, but also the labels of the edges for creating the random walks.

Wikipedia2Vec~\citep{Yamada:2016:JLE} is a skip-gram-based approach that applies graph embeddings to Wikipedia, creating embeddings that jointly capture link structure and text. 
The Wikipedia knowledge graph is a natural resource for training graph embeddings, considering that it represents entities in a graph of interlinked Wikipedia pages and their text. The method proposed in~\citep{Yamada:2016:JLE} embeds words and entities in the same vector space using word and graph contexts. 
The word-word context is modeled using the Word2Vec approach, entity-entity context considers neighboring entities in the link graph, and word-entity context takes the words in the context of the anchor text that links to an entity. 
The authors of Wikipedia2Vec demonstrate performance improvements on various NLP tasks, although they did not consider entity retrieval in their work.

\newreview{
Recent transformer-based methods utilize transformers to create vector representations of entities.
% have recently also be utized to create vector representations of entities. 
KGT5 \citep{Saxena:2022:sequence} treats link prediction as a sequence to sequence (seq2seq) task using a T5 architecture. It uses the prediction of sequences in the form of 
``predict tail:
\texttt{subject mention} $\mid$ \texttt{relation mention} $\mid$ '', and is trained on facts in the KG with the objective of generating the true answer using teacher forcing.  The method achieves competitive scores on link prediction on large datasets like Wikidata5M. 
This work is improved upon in KGT5-context \citep{Kochsiek:2023:friendly}, by adding extra context in the sequences, yielding even higher scores on link prediction on Wikidata5M. 
\cite{Li:2024:mocokgc} introduces MoCoKGC, which utilizes three different transformer-based encoders to create its embedding. It separately trains an entity-relation encoder, an entity encoder, and a momentum-entity encoder, which provides more negative samples and allows the gradual updating of entity encodings. 
%To our knowledge, it is the current state of the art on link prediction on Wikidata5M. 
In this work, we study three widely used and competitive KG embedding models with reasonable computational demands: ComplEx, RDF2Vec, and Wikipedia2Vec.}

\subsection{Entity Linking}\label{sec:elr}

Methods incorporating entity information in NLP or information retrieval rely on entity linkers to identify entity mentions in the query or document.
Entity linking refers to the process of detecting all possible mentions of entities and linking them to the corresponding identifier in a certain knowledge base~\citep{balog2018entity}. In this study, we employ the following five entity linking methods to annotate queries:

\new{In \citep{Hasibi:2015:ELQ}, the difference between entity linking in queries compared to full texts is discussed. Entity linking for queries is divided into two different tasks: The first is semantic mapping, which is finding a list of ranked entities similar to the queries. The second is interpretation finding, which finds sets of linked entities, representing possible interpretations, which can then be used for machine-understanding of the query. This work then also discusses evaluation methods for both tasks.  }

% \todo{Mention entity linking in queries and annotating multiple entities vs. one entity for a mention based on \citep{Hasibi:2015:ELQ, Hasibi:2017:ELQ}, because we mention in the intro and build on it in the method and evaluation sections.} 

TagMe \citep{ferragina2010tagme} is an on-the-fly entity linker specializing in short texts. It works in a three-step pipeline, parsing the query to make a candidate list of entities, then disambiguating. Afterward, it prunes entities with a low link probability/coherence to the other candidates. 
The REL entity linker~\citep{vanHulst:2020:rel, Joko:2022:PEC} is a popular open-source scalable entity linking toolkit~\citep{Kamphuis:2022:REBEL} for annotating various types of texts (e.g., documents and conversations) with entities. REL detects mentions using Flair, an NLP library that supports Named Entity Recognition (NER). REL performs candidate selection based on Wiki\-pe\-dia2Vec embeddings, and entity disambiguation based on latent relations between entity mentions in the text. 

A number of entity linkers are designed for annotating queries. Nordlys \citep{Hasibi:2017:NTE} uses the method created by \citep{Hasibi:2015:ELQ}, which employs a learning-to-rank (LTR) model with various textual and semantic similarity features. 
SMAPH \citep{Cornolti:2018:SMAPH} utilizes a so-called piggybacking approach that uses information from a web search engine to link entities. It does this by first using the text, which needs to be linked as a query, and building a set of candidate entities, which are then filtered by linking the candidate entities to the terms in the input query.  
ELQ \citep{li:2020:efficient} is a fast end-to-end entity linking model specialized for questions using bi-encoders. It performs both mention detection and entity linking in one pass. Its architecture encodes queries and entities, then scores candidate entities through the inner product with the entity vectors.

\subsection{Entity Retrieval}\label{sec:er}

Knowledge Graphs like Wikipedia enrich the representation of entities by modeling the relations between them. Methods for ad-hoc document retrieval, such as BM25, have been applied successfully to retrieve entity information from knowledge graphs. However, since
knowledge bases are semi-structured resources, this structural
information may be used as well, for example, by viewing entities as
fielded documents extracted from the knowledge graph. \newreview{BM25F~\citep{robertson2009probabilistic} is a fielded retrieval model}, where term frequencies between different fields in documents are normalized to the length of each field. Another effective model for entity retrieval is the
Fielded Sequential Dependence Model (FSDM) \citep{zhiltsov2015fielded},
which estimates the probability of relevance using information from
single terms and bi-grams, normalized per field.

Linking entities mentioned in the query to the knowledge graph allows relationships encoded in the knowledge graph to improve the estimation of the relevance of candidate entities. Previous
work has shown that entity linking can indeed help increase
effectiveness of entity retrieval. In \citep{hasibi2016exploiting}, for
example, entity retrieval has been combined with entity linking to
improve retrieval effectiveness over state-of-the-art methods
like FSDM.

% Our research uses the \textsc{TagMe} entity linker
% \citep{ferragina2010TagMe} because it is especially suited to annotate
% short and poorly composed text like the queries we need to link
% to. \textsc{TagMe} adds Wikipedia hyperlinks to parts of the text,
% together with a confidence score.

\citet{liu2019explore} is one of the early works applying graph embeddings to entity retrieval, which demonstrates consistent, albeit modest, improvements.
KEWER \citep{Nikolaev:2020:joint} is another work that introduces a method for retrieving entities based on graph embeddings. It learns embeddings for entities and words based on TransE and annotates queries using SMAPH to re-rank entities, which improved on the previous state of the art. Similarly, \citet{Gerritse:2020:GEEER} first annotate queries using TagMe, and then use Wikipedia2Vec embeddings to compute the similarity between queries and documents. This method also improves on the previous state of the art.

Following the popularity of transformer-based methods, various works have introduced combinations of transformers with graph embeddings. 
In \citep{Oza:2023:Entity}, the work of \citet{Gerritse:2020:GEEER} is extended to include transformer-based entity embeddings. 
\citet{daza:2021:inductive} introduces a BERT architecture combined with TransE graph embeddings to re-rank entities. After encoding the queries and entities, it uses their similarity as a query score.
In~\citep{Gerritse:2022:Entity}, a cross-encoder method is introduced, based on E-BERT~\citep{poerner:2020:ebert}. It introduces entity tokens in the input layer, of which the encodings are based on Wikipedia2Vec embeddings. 
In \citep{tran:2022:dense}, the BERT encoding of a query is combined with the Wikipedia2Vec embeddings of the annotated entities, which are aggregated using clusters. 
Last, \citet{chatterjee:2022:berter} construct a method for entities without entity embeddings, \newreview{by identifying the most relevant top-level sections from a Wikipedia page, depending on the query. }
 These sections are then used to train a BERT model to represent the entities, which, in turn, are used as features in an LTR model for entity retrieval. While existing approaches typically examine only a single graph embedding and entity linking method, this work explores the effect of various graph embeddings and entity linkers on entity retrieval.

% Very recent work has applied TransE graph embeddings to the problem
% of entity retrieval, and shown consistent but small improvements
% \citep{liu2019explore}. However, TransE graph embeddings are not a
% good choice if the graph has 1-to-many, transitive or symmetric
% relations, which is the case in knowledge graphs
% \citep{paulheimpresentation}. In our research, we also look into
% improving entity retrieval using graph embeddings, but use the
% Wikipedia2Vec representation to address these shortcomings.

\section{Embedding Based Entity Retrieval}
\label{sec:eer}

In this section, we describe the graph embedding and entity retrieval approaches used in this paper. We utilize three different graph embedding methods that are representative of the three types of graph embeddings. The first is Wikipedia2Vec, which combines graph-based and text-based structures and learns embeddings using skip-gram algorithms. Next is RDF2Vec, which is based on random walks. Finally, we use ComplEx, which represents transition-based graph embeddings. 
We conclude by describing the methodology of our graph embedding based on the re-ranking algorithm.

\subsection{Wikipedia2Vec}
\label{sec:eer:embeddings}

Taking a knowledge graph as the input, Wikipedia2Vec~\citep{Yamada:2016:JLE,Yamada:2018:WOT} extends the skip-gram variant of Word2Vec~\citep{Mikolov:2013:DRW,Mikolov:2013:EEW} and learns word and entity embeddings jointly for the Wikipedia knowledge graph. The objective function of this model is composed of three components.
The first component infers optimal embeddings for words $W$ in the corpus. Given a sequence of words $w_1 w_2 ... w_T$ and a context window of size $c$, the word-based objective function is:
\begin{equation}\label{eq:skipgram}
	\mathcal{L}_w = \sum_{t = 1}^T  \sum_{-c \leq j \leq c, j \neq 0}
	\log \frac{\exp(\mathbf{V}_{w_t}^\intercal \mathbf{U}_{w_{t+j}})}{\sum_{w\in W}
		\exp(\mathbf{V}_{w_t}^\intercal \mathbf{U}_{w})}, 
\end{equation}  
where matrices $\textbf{U}$ and $\textbf{V}$ represent the input and output vector representations, deriving the final embeddings from matrix $\textbf{V}$.

The two other components of the objective function take the knowledge graph into account. The first considers a link-based measure estimated from the knowledge graph (i.e., Wikipedia). This measure captures the relatedness between entities in the knowledge graph, based on the similarity between their incoming links:
\begin{equation}
	\mathcal{L}_e = \sum_{e_i\in \mathcal{E}} \sum_{e_o \in C_{e_i}, e_i \neq e_o}
	\log \frac{\exp(\mathbf{V}_{e_i}^\intercal \mathbf{U}_{e_o})}{\sum_{e\in
			\mathcal{E}}\exp(\mathbf{V}_{e_i}^\intercal \mathbf{U}_{e})},
\end{equation} 
where $C_e$ denotes entities linked to an entity $e$, and $\mathcal{E}$ represents all entities in the knowledge graph.
The last addition to the objective function places similar entities
and words near each other by considering the context of the anchor
text. The intuition is the same as in Word2Vec, but here,
words in the vicinity of the anchor text are used to predict the entity.
%entity has to predict the words in the anchor text.
Considering a knowledge graph with hyperlinks $A$ and an entity $e$, the goal is to predict the context words of the entity:
\begin{equation}
	\mathcal{L}_a = \sum_{e_i \in A} \sum_{w_o \in a(e_i)}
	\log\frac{\exp(\mathbf{V}_{e_i}^\intercal \mathbf{U}_{w_o})}{\sum_{w\in W}\exp(\mathbf{V}_{e_i}^\intercal \mathbf{U}_{w})}, 
\end{equation} 
where $a(e)$ gives the previous and next $c$ words of the referent entity $e$.

These three components (word context, link structure, and anchor context)
are then combined linearly into the following objective function:
% $\mathcal{L} = \mathcal{L}_w + \mathcal{L}_e + \mathcal{L}_a$.
\begin{equation}
	\label{eq:obj}
	\mathcal{L} = \mathcal{L}_w + \mathcal{L}_e + \mathcal{L}_a.
\end{equation}

\subsection{RDF2Vec}
RDF2Vec~\citep{Ristoski:2016:rdf2vec} is a graph embedding method that takes a collection of triples represented in the Resource Description Framework (RDF) format and generates entity-relation sentences using random walks throughout the graph. These sentences are then used to compute Word2Vec-based embeddings.  
Suppose we have an RDF graph $G = (R, \mathcal{E})$ where $R$ is a set of relations and $\mathcal{E}$ is a set of entities. 
RDF2Vec generates all paths $P_e$ in the RDF of depth $d$ for all entities $e\in\mathcal{E}$. These walks are generated using a breadth-first algorithm.  
%
% Changed to notation later introduced: 
\newreview{
First, for a starting entity $e_s$, the algorithm explores the direct outgoing relations $R(e_s)$. Of these edges, a path $e_s \rightarrow r_{i}$ is randomly selected, where $r_{i} \in R(e_s)$. Then, for each previously explored node, the algorithm will visit the connected vertices. This generates a path $e_s \rightarrow r_{i} \rightarrow e_{i}$. This will continue until $d$ iterations are reached, in which $d$ is a hyperparameter. }
After this, all paths $\cup_{e\in\mathcal{E}}P_e$ are entered as sentences into the SkipGram model, as is seen in equation \ref{eq:skipgram}. This will result in an embedding for all $e \in \mathcal{E}$, thus leading to an embedding space for $G$. 

\subsection{ComplEx}

%$h,t \in \mathbf{C}^d$, $r \in \mathbf{C}^d$, $Re(h^{t}diag(r)t)$
\newreview{
The TransE triple-based graph embeddings have been introduced in \citep{bordes2013translating}. The intuition behind TransE is to construct embeddings for head-relation-tail triples. Given triples $<e_h, r, e_t>$, where $e_h,e_t \in \mathcal{E}$, $r \in R$, it minimalizes the vectors $ \overrightarrow{e_h}, \overrightarrow{r}, \overrightarrow{e_t} \in \mathbb{R}^d$ with respect to the function  $ s(e_h,r,e_t) =  || \overrightarrow{e_h} + \overrightarrow{r} - \overrightarrow{e_h}||$. This results in an embedding space where the tail of the triple can be found using vector addition, i.e., $\overrightarrow{e_h} + \overrightarrow{r} = \overrightarrow{e_t}$.} However, TransE cannot embed one-to-many and many-to-many relations properly. Multiple algorithms expanding on TransE have been introduced, one of which is ComplEx~\citep{Trouillon:2016:Complex}, where entities are embedded in the complex field. 
 The intuition for utilizing complex vectors is to capture the many anti-symmetric relations in knowledge graphs more effectively than TransE, and this makes the computations arguably simpler since they use only the Hermitian dot product. %\newreview{The reason why we pick specifically ComplEx, is due to the work in \citep{Ruffinelli:2020:you}. This work shows that, when normalizing for similar setup and training, methods like TransE, ComplEx and DistMult have very similar performances to each other, when doing intensive parameter tuning. We chose ComplEx out of all the methods compared in \citep{Ruffinelli:2020:you}, because the authors have done parameter tuning of ComplEx on Wikipedia5M, which is a very similar dataset to the one used in this paper, with the intention of representing this class of methods as fairly as possible.  }
 
 %, while retaining the benefits of the dot product. 
 ComplEx first initializes random embeddings $\overrightarrow{e_h}, \overrightarrow{r}, \overrightarrow{e_t} \in \mathbb{C}^d$ for $e_h,e_t \in \mathcal{E}$, $r \in R$. Then, for all training triples, the following scoring function $S(\overrightarrow{e_h},\overrightarrow{r},\overrightarrow{e_t})$ is minimized: 
\begin{equation*}
	S(e_h,r,e_t) = \operatorname{Re}(\overrightarrow{e_h}^{T}\operatorname{diag}(\overrightarrow{r})\overrightarrow{e_t})
\end{equation*}
\new{where $\operatorname{Re()}$ is the function that maps a complex vector to its real part, and $\operatorname{diag()}$ the function that maps a vector of size $d$ to a matrix with dimensions $d\times d$, that has the input vector as diagonal and all other elements are 0. }

% \todo{Please explain what diag(r) and Re() functions mean here.}

We use the setup for sampling and loss computation described in \citep{Ruffinelli:2020:you}.
During training time, both positive and negative examples are presented, where the negative samples are obtained by randomly perturbing one of the head, relation, or tail entities.
The loss over these positive and negative samples is computed as follows: \newreview{first, apply the sigmoid function on $S(e_h,r,e_t)$, then compute cross-entropy loss over the resulting value and the label of that triple. }

%
%Let R and E be a set of relations and entities present in the KB. We want to recover the matrices of scores $X_r$, for all the relations $r \in R$. Given two entities $s$ and $o \in E$, the log-odd of the probability that the fact $r(s,o)$ is true is:
%
%\begin{equation}
%	P(Y_{rso} = 1)  = \sigma(\phi(r,s,o;\Theta))
%\end{equation}
%where $\phi$ is a scoring function that is typically based on a factorization of the observed relations and $\Theta$ denotes the parameters of the corresponding model. While X as a whole is unknown, we assume that we observe a set of true and false facts $\{Y_{rso}\}_{r(s,o)\in \Omega} \in \{-1,1\}^{|\Omega|}$, corresponding to the partially observed adjacency matrices of different relations, where $\Omega$ is the set of observed triples. The goal is to find the probability of $Yr's'o'$ being true or false for a set of targeted unobserved triples.
%
%Depending on the scoring function $\phi$ used to predict entries in the tensor X, we obtain different models. 
%
%The scoring function is: 
%\begin{equation}
%	\phi(r,s,o;\theta)= Re(<w_r, e_s, e_o>)
%\end{equation}
%where $w_r \in \mathbf{C}^K$ is a complex vector. 

\subsection{Re-ranking Entities}
\label{sec:eer:er}

In this section, we discuss the method of using these
graph embeddings in the setting of entity retrieval. 
We propose a two-stage ranking model, where we first produce a ranking of candidate entities using a high-performing baseline entity retrieval model (see Section \ref{sec:er}), and then use the graph embeddings to reorder these entities based on their similarity to the entities mentioned in the query, as measured in the derived graph embedding space. 

Following the related work discussed in Section \ref{sec:elr}, we use the selected entity linkers to identify the entities mentioned in
the query. Given input query $q$, we obtain a set of linked entities $E_q$ and a confidence score $s(e_q)$ for each entity $e_q\in E_q$, which represents the strength of the relationship between the query and the linked entity. We then compute an embedding-based score for the linked entities of query $q$ and entity $e$:
\begin{equation}\label{eq:score}
	F(e, E_q) = \sum_{e_q\in E_q} s(e_q) \cdot \operatorname{cos}(\overrightarrow{e}, \overrightarrow{e_q}),
\end{equation}
where $\overrightarrow{e}, \overrightarrow{e_q}$ denote the embeddings vectors for entities $e$ and $e_q$, respectively.
\begin{figure}[t]
    \centering
    \includegraphics[width=0.7\linewidth]{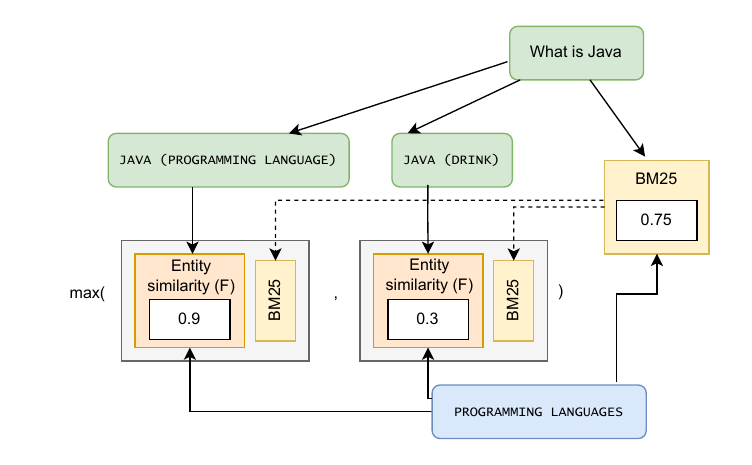}
    \caption{\newreview{A diagram illustrating our re-ranking approach for an example query with multiple entity linking interpretations. The re-ranking method computes the similarity score (Eq.~\ref{eq:score}) between target entity (Programming Language) and linked entities in each query interpretation. These scores will then be combined with the score of the initial retriever (BM25 in this example). The maximum of these scores is considered as the final re-ranking score.}}%. and. The method we use for reranking, in this example BM25, will be computed for the query-entity pair, and then the score based on the similarity between the target entity (Programming Language) and each scenario (Java Programming language, and Java Drink), will be computed. Then, the maximum of these scores will be picked as a final representation. }
    \label{fig:scenarios}
\end{figure}

The rationale for the scoring in Equation~\eqref{eq:score} is the hypothesis that relevant entities for a given query are situated close (in graph embedding space) to the query entities identified by the entity linker.
\newreview{Consider, for example, the query \emph{``Who is the daughter of Bill
	Clinton married to''}, which is linked to three entities by entity linker:
\textsc{Bill Clinton}, \textsc{Daughter}, and \textsc{Same-sex marriage}  with a confidence of scores of $0.66$, $0.13$, and  $0.21$, respectively; i.e., $E_q = \{$\textsc{Bill Clinton}, \textsc{Daughter}, \textsc{Same-sex marriage}$\}$.
The relevant entities for this query (according to the DBpedia-Entity V2 collection) are \textsc{Chelsea Clinton}, who is Bill Clinton’s daughter, and \textsc{Clinton Family}. 
%\textsc{Daughter}
% with a confidence of $0.13$, and \textsc{Same-sex marriage} with a
% confidence score of $0.21$. 
Highly ranked entities thus exhibit strong similarity to these linked entities, with similarity to \textsc{Bill Clinton} contributing more to the score than similarity to \textsc{Daughter} or \textsc{Same-sex marriage}, due to the higher confidence score associated with \textsc{Bill Clinton}.
Given their close semantic and relational ties, it is reasonable to expect relevant entities to be located near the linked entities in the embedding space, which confirms our intuition.
}
% The relevant entities for this query (according to the DBpedia-Entity
% V2 collection) are \textsc{Chelsea
% 	Clinton}, who is Bill Clinton's daughter, and \textsc{Clinton
% 	Family}. \newreview{Since these entities are closely related to each other, we can reasonably expect these entities to be close in the embedding space to the linked entities}, confirming our intuition. 

To produce our final score, we interpolate the embedding-based score
computed using Equation~\eqref{eq:score} with the score of the
baseline entity retrieval model $\mathit{score}_{\mathit{other}}$ used to produce the candidate
entities in stage one: 
\begin{align}
	\label{eq:lincomb}
	&\mathit{score}_{\mathit{total}}(e,q) = (1-\lambda) \cdot \mathit{score}_{\mathit{other}} (e, q) + \lambda \cdot F(e,E_q)
	& \lambda \in [0,1].
\end{align}
%

% In case we have multiple linked entity scenarios $E_1(Q), E_2(Q), \dots, E_n(Q)$ for a query, we will compute the score using the maximum of all scenarios, e.g. $E(Q) = \max_{i \in 1,\dots, n} E_i(Q)$.

When considering entity linking methods that return multiple interpretations of the query (i.e., multiple sets of $E_q$) , we  compute Equation~\eqref{eq:lincomb} for every entity linking interpretation $E^i_q$ and choose the interpretation that generates the highest score:
\begin{align}
    \label{eq:maxcomb}
    	&\mathit{score}_{\mathit{total}}(e,q) = \max_{E^i_q} \Big(  (1-\lambda) \cdot \mathit{score}_{\mathit{other}} (e, q) + \lambda \cdot F(e, E^i_q) \Big).
\end{align}

\newreview{Equation \ref{eq:maxcomb} extends Equation \ref{eq:lincomb} in the following way: It first computes $\mathit{score}_{\mathit{total}}(e,q)$ for every entity linking interpretation of the query and then chooses the highest score as the final score. 
We select the highest similarity score, rather than other aggregation methods, because a specific entity is typically relevant to only one interpretation of a query. For example, consider the query \emph{``What is Java?''}, which has three entity linking interpretations: $E_q^1 = \{ \textsc{Java (island)} \}$, $E_q^2 = \{ \textsc{Java (drink)} \}$, and $E_q^3 = \{ \textsc{Java (programming language)} \}$. The similarity between the entity \textsc{Programming Language} and the third interpretation $E_q^3 = \{ \textsc{Java (programming language)} \}$ would be high, while its similarity to the other two interpretations would be low. Since Programming Language is highly relevant to one specific sense of the query, taking the maximum score rather than an average is more appropriate.  A visual representation of this can be seen in Figure \ref{fig:scenarios}.}

% We pick the highest score, instead of a different aggregation method, because a specific entity is likely to only be relevant to a single scenario: Say that the query `What is Java' gets linked to the island, the coffee bean, and the programming language, the similarity between the page `Programming Language' and the Java programming language would be high, whereas it would be low to the other two scenarios. Since the entity `Programming Language' should be highly relevant to this query, taking only the maximum score instead of the average is preferable. A visual representation of this can be seen in Figure \ref{fig:scenarios}.}

% When extending this to entity-linking methods that consider multiple interpretations of the query, we combine the scores by picking the interpretation with the highest score. The intuition behind this is given a scenario $E_i(Q) \in E(Q)$, which scores high with an entity $e$; this scenario is likely the most representative of this query and entity. This method will likely give a better representation than aggregating my using a mean; when an entity scores high in one scenario but low in all others, the resulting mean will be low, which is not a fair representation.

% Thus when having multiple scenarios $E_i(Q) \in E(Q)$, the final score will be:

% \begin{align}
%     \label{eq:maxcomb}
%     	&\mathit{score}_{\mathit{scenario}}(E,Q) = \max_{E_i \in E} \mathit{score}_{\mathit{total}}(E_i,Q)
% \end{align}
\section{Experimental Setup}
\label{sec:exp}

%This section presents our experimental setup, including the test collection (\S\ref{sec:exp:coll}), embedding training (\S\ref{sec:exp:training}), and parameter settings (\S\ref{sec:exp:param}).

\subsection{Test collection}
\label{sec:exp:coll}
In our experiments, we use the standard entity retrieval collection, DBpedia-Entity V2 \citep{Hasibi:2017:DVT}. The collection consists of 467
queries and relevance assessments for 49280 query-entity pairs, where the entities are drawn from the DBpedia 2015-10 dump. The relevance assessments are graded values of 2, 1, and 0 for highly relevant, relevant, and non-relevant entities, respectively. The queries are categorized into 4 different groups: \textbf{SemSearch ES}  consisting
of short and ambiguous keyword queries (e.g.,\emph{``Nokia E73''}), \textbf{INEX-LD} containing IR-Style keyword queries (e.g., \emph{``guitar chord minor''}), \textbf{ListSearch} consisting of queries seeking for a list of entities (e.g., \emph{``States that border Oklahoma''}), and \textbf{QALD-2} containing entity-bearing natural language queries (e.g., \emph{``Which country does the creator of Miffy come from''}). Following the baseline runs curated with the DBpedia-Entity V2 collection, \newreview{we use the stopped version of queries provided by the dateset maintainers, where stop patterns like ``which'' and ``who'' are removed from the queries.}

\newreview{DBpedia-Entity V2 comprises queries from six benchmark evaluation campaigns, providing a diverse and heterogeneous set of queries. 
We point the reader to~\cite{Oza:2023:Entity}, which shows that our entity ranking method also generalizes to the TREC CAR dataset~\citep{dietz:2019:CAR} expanded with automatic entity annotations.}

%For generalization to the TREC CAR dataset~\citep{dietz:2019:CAR} with automatic entity annotations, we refer readers to~\cite{Oza:2023:Entity}, which demonstrates that our entity ranking method generalizes well to this dataset.

%Each query-entity pair gains a relevance assessment of highly relevant (2), relevant (1) or irrelevant (0). 
%It contains queries of four query types:
%
%\begin{itemize}
%	\item SemSearch ES: these are short queries, often looking for one specific entity like \emph{``Nokia E73.''}
%	\item INEX-LD: these are keyword queries, for example \emph{``guitar chord minor.''}
%	\item ListSearch: a combination of SemSearch LS, INEX-XER, and the TREC 2009 entity trec. These are queries looking for a certain list, like \emph{``States that border Oklahoma.''}
%	\item QALD-2: these are natural language questions, like \emph{``Which country does the creator of Miffy come from?''}.
%\end{itemize}
%In the usage of these queries, we use the stopped version, also provided in DBpedia-Entity V2. 
%This removes stopwords like `which' or `who' in queries.  \\ \\
%
\subsection{Entity Annotation}
\label{sec:exp:linking}

We use two sets of human-annotated queries as ground truth, Webis annotations provided by \citep{Kasturia:2022:query}, and Radboud annotations, a set created in this work.

\paragraph{Webis annotations.}
The annotations created in \citep{Kasturia:2022:query} present multiple scenario entity linking interpretations of the queries.  For example, a query like \emph{``java''} may have the island, the coffee, or the programming language as relevant documents, resulting in a set of interpretations where each interpretation relates to a different set of entities. Following \citep{Hasibi:2015:ELQ}, Webis annotations include only named entities (NEs).
% These different scenarios in the annotations are a valuable resource, \todo{as all entity linkers used in this research only provide one entity for a mention, even for ambiguous entities. } \fa{Are you sure? Nordlys provides multiple annotations}
%However, since this dataset only considers named entities and ignores concepts (e.g., \textsc{filmmaker}, \textsc{daughter}), we created a new dataset to enable us to investigate the importance of annotating concepts.

\paragraph{Radboud annotations.}
To compare the influence of concepts to that of named entities (NE), we construct a new set of annotations of DBpedia-Entity queries, which we refer to as Radboud annotations. We ask expert annotators to annotate all queries with any entity that would help solve the information needs of the queries. This leads to annotations of \emph{Named entities},  such as \textsc{Nokia} and \emph{Concepts} such as \textsc{movie producer} in query \emph{``who produced the film the ritual.''} 
% \begin{itemize}
% 	\item Named entities, such as `Nokia'
% 	\item Concepts, such as `books' or `movies'
% 	\item Job descriptions, such as `who produced the film...' get annotated to `movie producer'
% \end{itemize}
Annotators use the INCEpTION annotation platform \citep{Klie:2018:Inception} to link mention spans to WikiData instances. We let one expert annotate the entire dataset and let $2$ additional users annotate $50$ randomly selected queries, similar to the setup in \citep{Kasturia:2022:query}.

% \begin{table}
        
% 	\centering
% 	\caption{Pairwise annotator agreement}
%              \small

% 	\begin{tabular}{ l ||@{~}l @{~}l  }
% 		\hline
% 		      & cohens kappa & f-score   \\
% 		\Xhline{2pt}
%             % span cohens kappa also doesn't make any sense tbh. 
%             % Span & 0.41 & 0.57\\
%             %  & 0.43& 0.69 \\
%             Wikidata ID &0.54& 0.51 \\ 
%             & 0.59& 0.57 \\
% 		\hline
% 	\end{tabular}
% 	\label{tbl:agreement}
	
% \end{table}

We compute the F-measure and Cohen's Kappa for agreement, following \citep{deleger:2012:building,cui:2022:openel}. Cohen's Kappa score is computed on the token level (i.e., each word in a query is treated as a separate data point), a common strategy for handling multiple annotations per query. We found Kappa scores of $0.54$ and $0.59$, and F-scores of $0.51$ and $0.57$, which is sufficient for agreement \citep{Cohen:1960:coefficient}. An explanation for not having higher scores is that annotators might interpret a query differently. Since only one interpretation is asked, these annotations will then not agree. For example, the query \emph{``sri lanka government gazette''} could be linked to the two entities \textsc{SriLanka} and \textsc{The Sri Lanka Gazette}, but also to the non-overlapping entities \textsc{Government of Sri Lanka} and \textsc{Government Gazette}, which have high similarity in the embedding space (and are thus likely to only have small differences when using the differences to these entity embeddings), but still have an overlap of 0. % On top of that, some queries have multiple interpretations, of which annotators might select a different option.

\subsection{Entity Linking}
We employ the following entity linking methods: TagMe \citep{ferragina2010tagme}, REL \citep{vanHulst:2020:rel}, Nordlys \citep{Hasibi:2017:NTE}, SMAPH \citep{Cornolti:2018:SMAPH}, and ELQ \citep{li:2020:efficient}. To make a fair comparison, we use the default settings of all tools available, using the API if available. 
\begin{itemize}
    \item \emph{TagME and REL:} We use their APIs with the default settings.
    \item \emph{Nordlys:} Following \citep{Nikolaev:2020:joint}, we use the Nordlys toolkit API with the Learning to Rank option.
    \item \emph{SMAPH:} Following \citep{Nikolaev:2020:joint}, we annotate using the \textit{d4science} API, with the Google API as search engine. 
    \item \emph{ELQ:} We utilize the example script as listed on the official GitHub repository, using the default settings and preprocessing, most importantly casting all input to lowercase and using the \textit{elq wiki large} model.
\end{itemize}
% \textbf{TagMe:} For TagMe, we annotated using the API, using the default settings. 
    
% \textbf{REL:} For REL, we annotated using the provided API with its default settings. 
% \textbf{Nordlys:} 
% Following \citep{Nikolaev:2020:joint}, we use the Nordlys toolkit with option LTR. using the API. \\
% \textbf{SMAPH:}
% Following \citep{Nikolaev:2020:joint}, we annotate using the \textit{d4science} API, with the Google API as search engine. \\
% \textbf{ELQ:} For ELQ, we utilized the example script as listed on the official GitHub repository, using the default settings and preprocessing, most importantly casting all input to lowercase and using the \textit{elq wiki large} model.\\

\subsection{Embedding Training}
\label{sec:exp:training}

\paragraph{Wikipedia2Vec.}
% We finetune Wikipedia2Vec using the same setup of   \citep{Gerritse:2020:GEEER}: setting the min-entity-count parameter to zero and using the link graph during training. We use the original version used in \citep{Gerritse:2020:GEEER}, where they used the Wikipedia 2019 dump and spent a substantial amount of effort on fixing missing entities by solving redirects or correcting errors in naming. To make the comparison fair to the other graph embedding methods, we also use the Wikipedia 2015-10 dump, the version on which DBpedia 2015-10 was based. 

%+++++++++++++++
Wikipedia2Vec provides pre-trained embeddings. % based on the Wikipedia 2018-04 dump. 
These embeddings, however, are not available for all
entities in Wikipedia; e.g., 25\% of the assessed entities in
DBpedia-Entity V2 collection have no pre-trained embedding. The
reasons for these missing embeddings are two-fold: (i) ``rare''
entities were excluded from the training data, and, (ii) entity
identifiers evolve over time, resulting in entity mismatches
with those in the DBpedia-Entity collection.

For training new graph embeddings, we used the Wikipedia 2019-07 dump, which is also compatible with recent entity linkers. We address the entity mismatch problem by identifying the entities that
have been renamed in the new Wikipedia dump. Some of these entities
were obtained using the redirect API of
Wikipedia.\footnote{\url{https://wikipedia.readthedocs.io/en/latest/}}
Others were found by matching the Wikipedia page IDs of the two
Wikipedia dumps. The page IDs of Wikipedia 2019-07 were available on
the Wikipedia website. For the dump where DBpedia-Entity is based on, however, these IDs are
not available anymore; we obtained them from the Nordlys
package~\cite{Hasibi:2017:NTE}.

To avoid excluding rare entities and generate embeddings for a wide
range of entities, we changed several Wikipedia2Vec settings. The
two settings that resulted in the highest coverage of entities are:
(i) minimum number of times an entity appears as a link in Wikipedia,
(ii) whether to include or exclude disambiguation
pages. Table~\ref{tbl:missingent} shows the effect of these settings
on the number of missing entities; specifically the number of entities
that are assessed in the DBpedia-Entity collection, but have missing
embeddings. We categorize these missing entities into two groups:
% three groups:
 %
\begin{itemize}
	%\item \emph{Disamb}: Entities which get a \emph{Disambiguation error} after using the API. 
	\item \emph{No-page}: Entities without any pages, \newreview{i.e., although they have an identifier, there is no actual Wikipedia page with that exact identifier name}. These entities were neither found by the Wikipedia redirect API nor could they be matched by their page IDs. 
	\item \emph{No-emb}: Entities that could be found by their identifiers, but were not included in the Wikipedia2Vec embeddings.
\end{itemize}

The first line in Table~\ref{tbl:missingent} corresponds to the
default setting of Wikipedia2Vec, which covers only 75\% of assessed
entities in the DBpedia-Entity collection. When considering all entities
in the knowledge graph, this setting discards an even larger number of
entities, which is not an ideal setup for entity ranking. By choosing
the right settings (the last line of Table~\ref{tbl:missingent}), we
increased the coverage of entities to 97.6\%.% These missing embeddings are mainly due to the entity mismatch between the two Wikipedia versions, which is inevitable.

To make the comparison fair to the other graph embedding methods, we also use the Wikipedia 2015-10 dump, the version on which DBpedia 2015-10 was based. 
 
% We trained two versions of embeddings: with and without link graph;
% i.e., using Eq.~\eqref{eq:obj} with and without the $\mathcal{L}_e$
% component.

\begin{table}[t]
	\centering
	\caption{Missing entities with different settings.}
	\begin{tabular}{ l || lll }
		\toprule
		\textbf{Settings} & \textbf{No-emb} & \textbf{No-page} & \textbf{Total} \\
		\hline
		min-entity-count = 5, disambiguation = False & 9640    & 608    & 10248 \\
		min-entity-count = 1, disambiguation = False & 1220    & 398    & 1618  \\
		min-entity-count = 1, disambiguation = True & 1220    & 377    & 1597  \\
		min-entity-count = 0,  disambiguation = False                  & \textbf{724} & 380 & 1104  \\
		min-entity-count = 0,  disambiguation = True  & \textbf{724} & \textbf{333} & \textbf{1057} \\
		\bottomrule
	\end{tabular}
	\label{tbl:missingent}
\end{table}
%+++++++++++++++++++

\paragraph{RDF2vec.}
RDF2Vec and ComplEx are trained using the DBpedia link graph as input. DBpedia consists of over 20 possible input files, some containing more relevant information for retrieval-oriented graph embeddings than others. 
For training RDF2vec and ComplEx, we use the following files from DBpedia:  

\begin{itemize}
	\item Disambiguations:  Links extracted from Wikipedia Disambiguation pages, \newreview{which are the Wikipedia pages that redirect a user when there are multiple entities with the same name.}
	\item Infobox properties: Information which was extracted from the Wikipedia Infoboxes.
	\item Instance types: Triples of the form object, RDF type, and class from the mapping-based extraction.
	\item Instance types transitive: Transitive RDF type-class based on the ontology.
	\item Mapping-based objects: Mapping-based statements with object values.
	\item Transitive redirects: Transitively resolved redirects between articles in Wikipedia.
	\item Pagelinks: Contains internal links between entities on DBpedia, created based on the internal links between Wikipedia pages.
\end{itemize}

In the original RDF2Vec paper, the files used for finetuning are as listed above, except for the `Pagelinks' file. Including this file increases the graph size from 11GB to 35GB, which in turn increases the training time and the final embedding output size. However, excluding this file results in missing a substantial number of entities, as seen in Table~\ref{tbl:missingentemb}. 
Since several triples in the Pagelinks files appear to include alternatives for semantically similar entities, incorporating these nodes and edges could potentially dilute the effect of fine-tuning. \newreview{We hypothesize that this occurs because entities seen sparsely during fine-tuning tend to have lower-quality representations in the embedding space. Distributing such an entity across multiple semantically similar entities may further degrade the quality of embeddings.} As a resolution, we include results with and without the Pagelinks file. 

 % We reason this could happen, because for an entity to be seen sparsely during finetuning, negatively impacts the quality of its location in the embedding space. Then, spreading this entity over multiple semantically similar entities, might deteriorate the quality even further.}

For finetuning RDF2Vec, we use the jRDF2VEC package \citep{Ristoski:2016:rdf2vec}. We finetune using the default settings of the package, which is \textit{walk depth} of 4, \textit{numberOfWalks} of 100 per entity, \textit{walkGenerationMode} as Random, \textit{Dimension size} of 200 and \textit{Number of Epochs} of 5. 

%java -jar jrdf2vec-1.1-SNAPSHOT.jar -onlyTraining -walkDirectory walks_1306/ -dimension 200

\paragraph{ComplEx.}
%CUDA_VISIBLE_DEVICES=1 python -m kge start dbpedia_small-complex_gpu.yaml 
For ComplEx, we use the same training files as for RDF2Vec, reusing the same setup with and without the Pagelinks file. We use the LibKGE \citep{Broscheit:2020:libkge} package and the same configuration as used with the Wikipedia5M dataset, which has similar properties to DBpedia. 

Table \ref{tbl:missingentemb} shows the number of missing entity embeddings per embedding type. This is after using the DBpedia redirect file to solve redirects. We can see that, even when using DBpedia 2015-10 and Wikipedia 2015-10, many files are still missing in the eventual embedding space, which is bound to hurt results for re-ranking.

\begin{table}
	\centering
        \caption{Number of missing entities with different settings of graph embeddings.
        }
	\begin{tabular}{l ||@{~}l} %{0.45\linewidth}
		\toprule
		\textbf{Embedding} &  \textbf{\#Missing entities}  \\
		\hline
            Wikipedia2Vec 2019 & 36326 \\
		Wikipedia2Vec 2015 &  14124  \\
		ComplEX &  25512  \\
		ComplEX Pagelinks & 58\\
		RDF2Vec & 31782  \\
		RDF2Vec Pagelinks & 75 \\
		\bottomrule
	\end{tabular}
	\label{tbl:missingentemb}
\end{table}

\subsection{Evaluation metrics} \label{ssec:eval}

\paragraph{Entity Linking.}
% Given that the entity linkers employed in this paper only yield one scenario,} 
Given that Webis annotations contain multiple entity linking scenarios and Radboud annotations contain only one scenario, we need an evaluation method to accommodate both formats. 
One could choose, for example, evaluating exclusively the best possible scenario, the average across all scenarios, or the union of all scenarios as a ground truth. However, using these strategies, comparing queries with a difference in the number of scenarios presents a challenging task.
We, therefore, use the lean evaluation metrics introduced by \citet{Hasibi:2015:ELQ}.

Suppose $\hat{I} = \{\hat{E}_1, ..., \hat{E}_m\}$ denote the query interpretations for the ground truth, and $I = \{E_1, ..., E_n\} $ the interpretation returned by the system. \newreview{Here, each $\hat{E}_i$, $E_i$ is a set of entities, and thus $\hat{I}, I$ are sets of sets of entities.} Let $\hat{E} = \bigcup \hat{E}_i$ and  $E= \bigcup E_i$. \newreview{We first define the precision and recall based on the different interpretations, which we call  $P_{int}$ and $R_{int}$: }
% Then we define  $P_{int}$, $R_{int}$ to be: 
%$$P_{int} = \frac{|\hat{I}\cap I|}{|I|}$$,
%$R_{int} = \frac{|\hat{I}\cap I|}{|\hat{I}|}$ 
%
\begin{align*}
P_{int} =&\left\{
\begin{aligned}
\frac{|\hat{I}\cap I|}{|I|} , &\quad I \neq \emptyset \\
1 , &\quad I = \emptyset, \hat{I} = \emptyset \\
0 , &\quad I = \emptyset, \hat{I} \neq \emptyset\\
\end{aligned}\right. 
&
R_{int} = &
\left\{\begin{aligned}
\frac{|\hat{I}\cap I|}{|\hat{I}|} , &\quad  \hat{I} \neq \emptyset\\
1  , & \quad  \hat{I} = \emptyset, I = \emptyset \\
0 , & \quad  \hat{I} = \emptyset, I \neq \emptyset \\
\end{aligned}\right. \\
\end{align*}
\newreview{We then define the precision and recall based on all the entities linked, which we refer to as  $P_{ent}$ and $R_{ent}$: }
 % $P_{ent} = \frac{|\hat{E}\cap E|}{|E|}$,
\begin{align*}
P_{ent} =&\left\{
\begin{aligned}
\frac{|\hat{E}\cap E|}{|E|} , &\quad E \neq \emptyset \\
1 , &\quad E = \emptyset, \hat{E} = \emptyset \\
0 , &\quad E = \emptyset, \hat{E} \neq \emptyset\\
\end{aligned}\right. 
&
R_{ent} = &
\left\{\begin{aligned}
\frac{|\hat{E}\cap E|}{|\hat{E}|} , &\quad  \hat{E} \neq \emptyset\\
1  , & \quad  \hat{E} = \emptyset, E = \emptyset \\
0 , & \quad  \hat{E} = \emptyset, E \neq \emptyset. \\
\end{aligned}\right. \\
\end{align*}
%
%$R_{int} = \frac{|\hat{E}\cap E|}{|\hat{E}|}$.\\ \\
Lean precision and recall are then the combination between these two scores, being: 
\begin{equation*}
    P_{lean} = \frac{P_{int} + P_{ent}}{2}, \quad  R_{lean} = \frac{R_{int} + R_{ent}}{2}.
\end{equation*}
%$$P_{lean} = \frac{P_{int} + P_{ent}}{2}, \quad  R_{lean} = \frac{R_{int} + R_{ent}}{2}$$.
%

When given only one scenario for both system annotations and ground truth, we see that $P = P_{lean} = P_{int} = P_{ent}$ and $R = R_{lean} = R_{int} = R_{ent}$. 
Thus, lean evaluation can be utilized for Webis annotations, which encompass multiple scenarios, as well as Radboud annotations, which do not include multiple scenarios. Lean evaluation can be interpreted as the standard precision and recall for the latter.

% Therefore using lean evaluation works for both Webis annotation, which includes multiple scenarios, and Radboud annotations, which do not include scenarios. 
\paragraph{Entity retrieval.}

We evaluate each combination of all methods and queries using the method used in \citep{Hasibi:2017:DVT}, which is the Normalized Discounted Cumulative Gain (NDCG) at ranks 10 and 100. We report on statistical significance for NDCG@10 and NDCG@100 using a two-sided t-test with  p-value $<$ 0.05.
 \section{Results and Analysis}
\label{sec:results}

\subsection{Entity Linking Results}

To compare all different entity linking methods, we compute precision, recall, and F-measure for all entity linkers in Table \ref{tbl:linkedent}, using lean evaluation described in Section~\ref{ssec:eval}.
The line `combined' indicates the score for the combination of all linked entities, which is the union of TagMe, REL, Nordlys, SMAPH, and ELQ. 
 With Webis as ground truth, Nordlys has the highest recall and precision. Using our concept annotations as ground truth, ELQ has the highest precision, and TagMe has the highest recall. As seen in Table \ref{tbl:linkedent}, there is a difference in what each of these entity linkers excels in, resulting in no single best method. 

\begin{table}
\label{tbl:el}
        \centering
	\caption{Performance of different entity linkers against Webis (only NE) and Radboud (both NEs and concepts) annotations.
 %.Named Entity (NE) scenarios and concepts + NE. 
Lean evaluation is used for multiple interpretations (Webis), and regular precision/recall is used for single interpretations (Radboud). `Combined' refers to the union of all entities returned by all linkers.}
	\begin{tabular}{ l ||c |ccc |ccc }
		\toprule
              & \textbf{\#Linked } & \multicolumn{3}{c|}{\textbf{Webis (NEs)}}&\multicolumn{3}{c}{\textbf{Radboud (Concepts+NEs)}} \\
		   & \textbf{entities} & P$_{lean}$  & R$_{lean}$    & F$_{lean}$    & P  & R  & F\\
		\hline
		TagMe & 1186 & 0.479 & 0.598 & 0.532&  0.704 & \textbf{0.814 }& \textbf{0.755} \\
		REL & 363 & 0.699 & 0.618 & 0.656 & 0.534 & 0.416 & 0.467\\
		Nordlys & 450 &\textbf{0.722} & \textbf{0.642} & \textbf{0.680} & 0.553 & 0.439 & 0.490\\
		SMAPH & 797 & 0.526 &0.550 & 0.538 &0.758  & 0.720 & 0.739 \\
		ELQ & 647 &0.606 & 0.587 & 0.597 & \textbf{0.787} & 0.701 & 0.741 \\
		Combined & 1795 & 0.437 & 0.686 & 0.534 &0.621 & 0.911 & 0.738  \\
		%Ground-truth Webis & 830 & - & - & - &0.605 & 0.552 & 0.577\\
		%Ground truth Concepts &891 & 0.55 &0.603 &0.575 & - & - & -    \\
		\bottomrule
	\end{tabular}
	\label{tbl:linkedent}	
\end{table}

\begin{table}[t]
    \centering
    \caption{Entity retrieval results on DBpedia-Entity V2 collection using different entity linkers and gold annotations. Wikipedia2Vec 2019 is used for re-ranking of BM25F-CA results. Superscripts denote statistically significant differences (better or worse) corresponding to the beginning letter of entity linkers' names, BM25F-CA, and Webis.}
    \begin{tabular}{l || l l}
    \toprule
         % & \multicolumn{2}{c}{\bfseries Total}\\
		%    \hline
    & \textbf{NDCG@10} & \textbf{NDCG@100} \\
       \hline
    BM25F-CA
    &  0.461 & 0.551 \\
    + Wikipedia2Vec (w/ TagMe) 
    & 0.484$^{\tiny b}$ & 0.570$^{\tiny b}$ \\
    + Wikipedia2Vec (w/ SMAPH)
    & 0.483$^{\tiny b}$ & 0.570$^{\tiny b}$ \\
    + Wikipedia2Vec (w/ Nordlys) 
    & 0.477$^{\tiny b}$ & 0.563$^{\tiny bts}$ \\
    + Wikipedia2Vec (w/ REL) 
    & 0.472$^{\tiny bts}$ & 0.561$^{\tiny bts}$ \\
    + Wikipedia2Vec (w/ ELQ)
    & \textbf{0.489}$^{\tiny bnr}$ & \textbf{0.573}$^{\tiny bnr}$ \\
    \hline
    + Wikipedia2Vec (w/ Combined)
    & 0.498$^{\tiny btsnre}$ & 0.58$^{\tiny btsnre}$ \\
    \hline
    + Wikipedia2Vec (w/ Webis)
    & 0.475$^{\tiny bea}$ & 0.563$^{\tiny btsea}$ \\
    + Wikipedia2Vec (w/ Radboud)
    & 0.492$^{\tiny bsnrw}$ & 0.577$^{\tiny bnrw}$ \\
    \bottomrule
    \end{tabular}
    \label{tbl:res_el}
\end{table}

\begin{table}[t]
\centering
	 \caption{Entity retrieval results on DBpedia-Entity V2 collection using different graph embeddings. Superscripts denote statistically significant differences (better or worse) corresponding to that line of the table.} 
	\label{tbl:res2}
	\begin{tabular}{l || ll }
		 \toprule
		& \textbf{NDCG@10} & \textbf{NDCG@100} \\
		 \hline
		\multicolumn{3}{l}{\emph{Base}}\\
		\hline
		Wikipedia2Vec 2019 
            & \textbf{0.262}$^{1}$ & \textbf{0.335}$^{1}$\\
		Wikipedia2Vec 2015 
            & 0.184$^{12}$ & 0.278$^{12}$ \\
		ComplEx  
            & 0.182$^{12}$ & 0.216$^{123}$  \\
		ComplEx Pagelinks  
            & 0.204$^{1234}$ & 0.255$^{1234}$
		\\
		RDF2Vec  
            & 0.188$^{125}$ & 0.23$^{12345}$\\
		RDF2Vec Pagelinks 
            & 0.195$^{12}$ & 0.262$^{12346}$ \\
		\hline
		\multicolumn{3}{l}{\emph{TagMe}}\\
		\hline
		BM25F-CA  
            & 0.461 & 0.551 \\
            + Wikipedia2Vec 2019  
            & 0.484$^{1}$ & 0.57$^{1}$\\
            + Wikipedia2Vec 2015  
            & 0.466$^{2}$ & 0.559$^{12}$\\
            + ComplEx   
            & 0.468$^{12}$ & 0.558$^{12}$\\
            + ComplEx Pagelinks   
            & 0.474$^{1234}$ & 0.564$^{1234}$ \\
            + RDF2Vec   
            & 0.461$^{245}$ & 0.554$^{235}$ \\
            + RDF2Vec Pagelinks   
            & 0.467$^{256}$ & 0.56$^{126}$ \\
		\hline
		\multicolumn{3}{l}{\emph{Radboud annotations}}\\
			\hline
   		BM25F-CA  
            & 0.461 & 0.551 \\
            + Wikipedia2Vec 2019  
            & \textbf{0.492}$^{1}$ & \textbf{0.576}$^{1}$ \\
            + Wikipedia2Vec 2015   
            & 0.473$^{12}$ & 0.564$^{12}$ \\
            + ComplEx   
            & 0.467$^{12}$ & 0.558$^{123}$ \\
            + ComplEx Pagelinks   
            & 0.476$^{124}$ & 0.564$^{124}$ \\
            + RDF2Vec   
            & 0.463$^{235}$ & 0.556$^{1235}$ \\
            + RDF2Vec Pagelinks   
            & 0.471$^{126}$ & 0.566$^{1246}$\\
%            \hline
%		\multicolumn{3}{l}{\emph{Transformer-based}}\\
%			\hline
%            MonoBERT & 0.508 & 0.586 \\
%            EM-BERT  & 0.541 & 0.604 \\
		\bottomrule
	\end{tabular}
 \label{tbl:res_emb}
\end{table} 

\subsection{Entity Retrieval Results}
\label{sec:results:overall}
Table~\ref{tbl:res_el} depicts entity retrieval results using different automatic entity linkers and human annotations; the breakdown of results by query type can be found in Table~\ref{tbl:res_el_full} of the Appendix. We use the Wikipedia2Vec embeddings from \citep{Gerritse:2020:GEEER} and perform re-ranking with BM25F-CA.
When using human annotations, Radboud annotations receive better results than the multiple scenarios used in the Webis annotations. Especially in the SemSearch, INEX, and ListSearch categories, we see an increase in both NDCG@10 and NDCG@100 using Radboud annotations. \newreview{Since the Radboud annotations focuses on adding concept annotations,} this indicates that concept entities are, in fact, an essential factor for entity retrieval using entity embeddings.

The entity retrieval methods using TagMe, SMAPH and ELQ as entity linkers obtain the best results, with no overall significant differences between each other. These three entity linkers obtain the highest F-measure when using the Radboud annotations as ground truth. This answers our first research question \textbf{RQ1}: Entity-linking methods with a high F-measure with respect to both concepts and named entities are best suited for the entity retrieval method discussed in this paper.

Table~\ref{tbl:res_emb} shows the results for different embedding algorithms using TagMe and Radboud annotations; the full table can be found in Table~\ref{tbl:res_emb_full} of the Appendix. In the first block, denoted as \textit{`base'}, we only rank using the entity similarity score per embedding type; i.e., using Equation~\ref{eq:score} only, \newreview{using TagMe as an entity linker}.
%In the next block, denoted as \textit{ `all'}, we take the union between the linked TagMe, SMAPH, Nordlys and Rel annotations. 
Next, we re-rank the BM25F-CA results, both using TagMe (as used in \citep{Gerritse:2020:GEEER}) and Radboud annotations. We see that Wikipedia2Vec embeddings outperform ComplEx and RDF2Vec in base form, but not in the base form where many entities are missing. We also see that RDF2Vec and ComplEx perform significantly better when using the Pagelink pages, indicating that they are essential when re-ranking with entity retrieval. Last, we see that Radboud annotations yield higher scores than the TagMe annotations.

\subsection{Embedding Analysis}

% For judging the quality of the different embedding types, we compute the query coherence score as defined in \citep{he2011exploring} for each query.
% % we use the formula for query coherence defined in \citep{he2011exploring}. 

% This score first views all entities relevant to the same query as a cluster, and then for those embeddings counts the number of pairs where the similarity between the two embeddings is higher than a certain threshold. Thus, queries where relevant entities are similar to eachother will have a higher score.

% Formally, given a document set $D$, the coherence score
% is computed as:
% %
% \begin{equation}
% 	Co(D) = \frac{\sum_{i\neq j \in {1, \dots, M} } \delta(d_i, d_j) }{\frac{1}{2} M(M-1)},
% \end{equation}
% %
% where $M$ is total number of documents and the $\delta$ function for
% each document pair $d_i$ and $d_j$ is defined as:

% \begin{equation}
% 	\delta(d_i,d_j) =   
% 	\begin{cases}
% 		1, & \text{if}\ sim(d_i,d_j)\geq \tau \\
% 		0, & \text{otherwise}.
% 	\end{cases}
% \end{equation}

To judge how suitable each graph embedding is for retrieval, we compare the embeddings of documents relevant to the same query, based on the cluster hypothesis \citep{rijsbergen1979information}.
This states that documents relevant to the same query should cluster together in a higher dimensional space. Here, we consider the entity embedding of a document as its representation.
Using this hypothesis, we compute each query's coherence score as defined in \citep{he2011exploring}, which measures the similarity between all pairs of documents relevant to the same query and returns the percentage of items with a similarity score higher than a
threshold. Formally, given a document set $D$, the coherence score
is computed as:
\begin{equation}
	Co(D) = \frac{\sum_{i\neq j \in {1, \dots, M} } \delta(d_i, d_j) }{\frac{1}{2} M(M-1)},
\end{equation}
where $M$ is total number of documents and the $\delta$ function for
each document pair $d_i$ and $d_j$ is defined as:
% This score computes the percentage of item pairs in a cluster that are close to each-other more than a certain threshold . Thus giving a higher score to clusters which are more coherent. Given a document set $D$ and a threshold $\tau$, we have:
%
\begin{equation}
	\delta(d_i,d_j) =   
	\begin{cases}
		1, & \text{if}\ sim(d_i,d_j)\geq \tau \\
		0, & \text{otherwise}.
	\end{cases}
\end{equation}

\begin{figure*}[t]
	%\shrink
	\centering
	\includegraphics[width=0.6\textwidth]{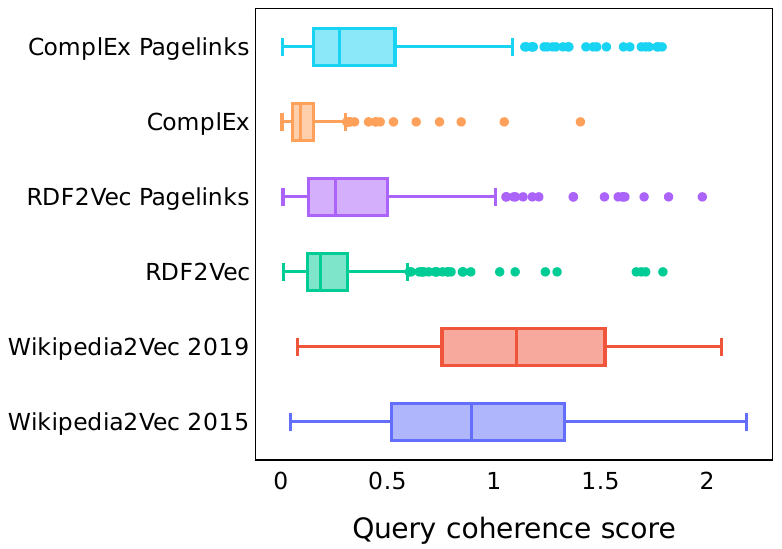}
	%	 \shrink
	\caption{Query coherence score of three different graph embedding algorithms. Higher coherence score is better. Wikipedia2Vec has higher coherence scores compared to  RDF2Vec and ComplEx .}
	\label{fig:boxplot}
\end{figure*}

In Figure~\ref{fig:boxplot}, we see coherence scores for the different entity embedding methods used in this paper, \newreview{with a threshold of $\tau = 0.7$. This threshold is selected via grid search as the highest value at which none of the box plots had a first quartile equal to 0.} For RDF2Vec and ComplEX, we include both versions with and without pagelinks. We compute the coherence score on the 295 queries with at least ten relevant entities. The higher the query coherence scores are, the better the distance between embeddings should be able to represent relevance for documents to the same query. We see that Wikipedia2Vec leads to the highest coherence scores. Interestingly enough, we see that the 2019 version of Wikipedia has a higher average coherence score than the 2015 version, even though more entities seem to be missing. A reason for this could be that the quality of Wikipedia has improved over the years, with more new pages and additional text added, resulting in a better coherence score. 
%Another reason could be that in \citep{Gerritse:2020:GEEER}, a great effort was made to find the redirects for missing pages. 
Besides, we see that for both ComplEX and RDF2Vec, the additional page links improve the coherence scores. 

We now answer our second research question \textbf{RQ2}: Wikipedia2Vec leads to the best results for the entity retrieval method discussed in this paper. However, it is essential to invest in solving the missing entities for optimal performance. For ComplEx and RDF2Vec, including a sufficiently large number of entities in the graph is considerably important.

\subsection{Wikipedia2Vec Embedding Analysis}
\begin{figure*}[t]
	\centering
	\begin{subfigure}[t]{0.49\textwidth}
		\includegraphics[width=\textwidth]{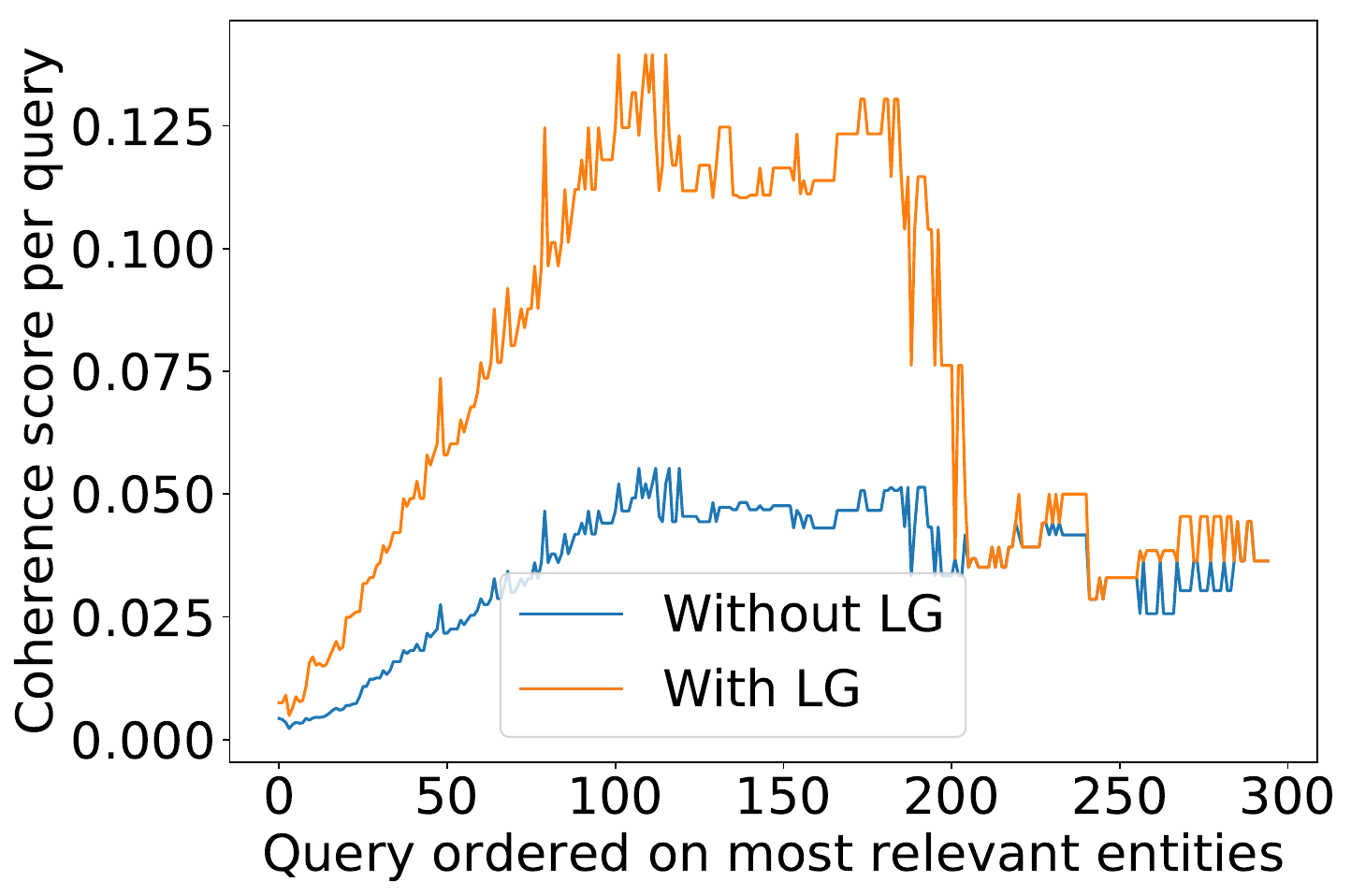}
		\caption{threshold $\tau=0.9$}
		\label{fig:coh:9}
	\end{subfigure}
	\begin{subfigure}[t]{0.49\textwidth}
		\includegraphics[width=\textwidth]{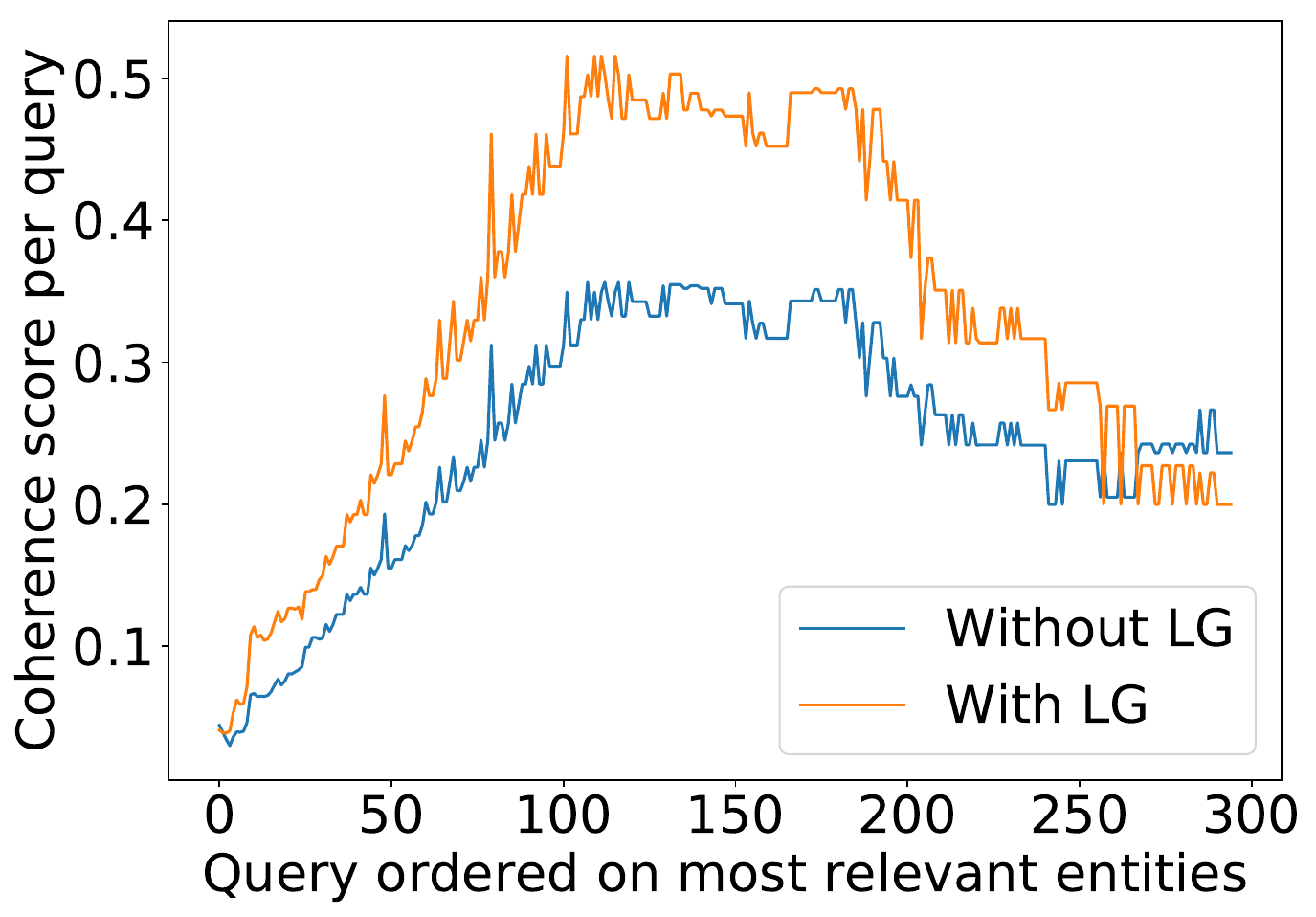}
		\caption{threshold $\tau=0.8$}
		\label{fig:coh:8}
	\end{subfigure}
	\caption{Coherence score of all relevant entities per query, computed for the Wikipedia2Vec embeddings without and with link graph. The queries are ordered by the number of their relevant entities in the x-axis.} %\newreview{Since the coherence score is displayed per query, the threshold was put higher than Figure \ref{fig:boxplot}, to highlight more detailed differences between these two embeddings.} }
	\label{fig:coh}
\end{figure*}
We empirically showed that Wikipedia2Vec graph
embeddings yield better performance compared to other embeddings. To analyze why Wikipedia2Vec graph embeddings are
beneficial for entity retrieval models, we conduct a set of experiments and investigate how the graph structure captured by Wikipedia2Vec embeddings improves effectiveness. Specifically, we trained two versions of Wikipedia2Vec embeddings: with and without link graph; i.e., using Eq.~\eqref{eq:obj} with and without the $\mathcal{L}_e$ component.

Figure~\ref{fig:coh} shows the coherence scores for all queries
in our collection. Each point represents the coherence score of all
relevant entities (according to the qrels) for a query. We considered
only queries with more than 10 relevant entities, ensuring the clusters were sufficiently large to yield meaningful scores.
% for clusters large enough to compute a meaningful score. 
Queries are sorted on the x-axis
by the number of relevant entities.
The plots clearly show that the coherence score for graph-based
entity embeddings are higher than for context-only ones. Based on these performance improvements, we conclude that adding the
graph structure results in embeddings that are more suitable for
entity-oriented tasks.

\begin{figure*}[t]
	\centering
	\begin{subfigure}[t]{0.48\textwidth}
		\includegraphics[width=0.98\linewidth]{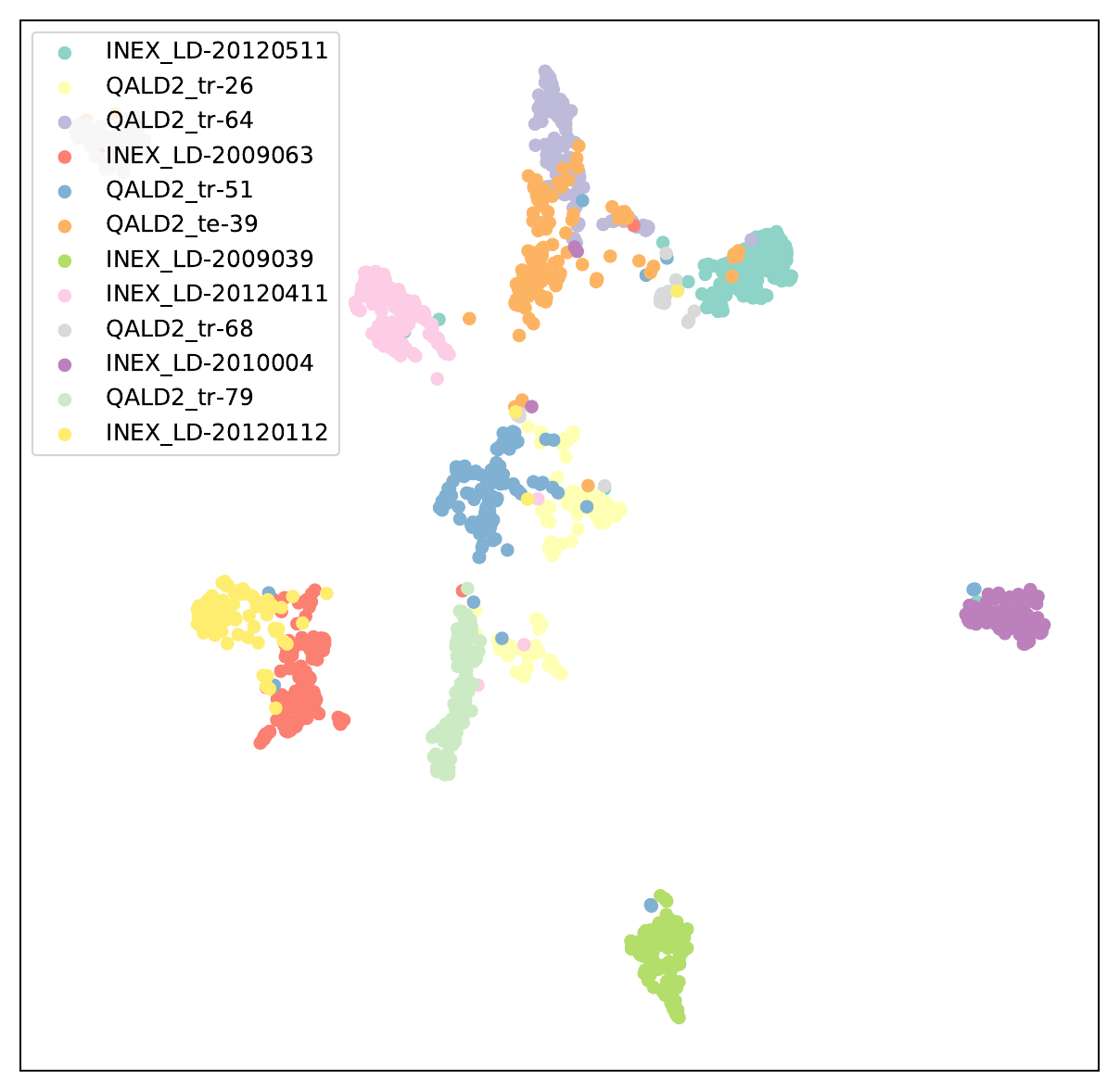}
		\caption{Embeddings with link graph}
		\label{fig:umap:lg}
	\end{subfigure}
	\begin{subfigure}[t]{0.48\textwidth}
		\includegraphics[width=\linewidth]{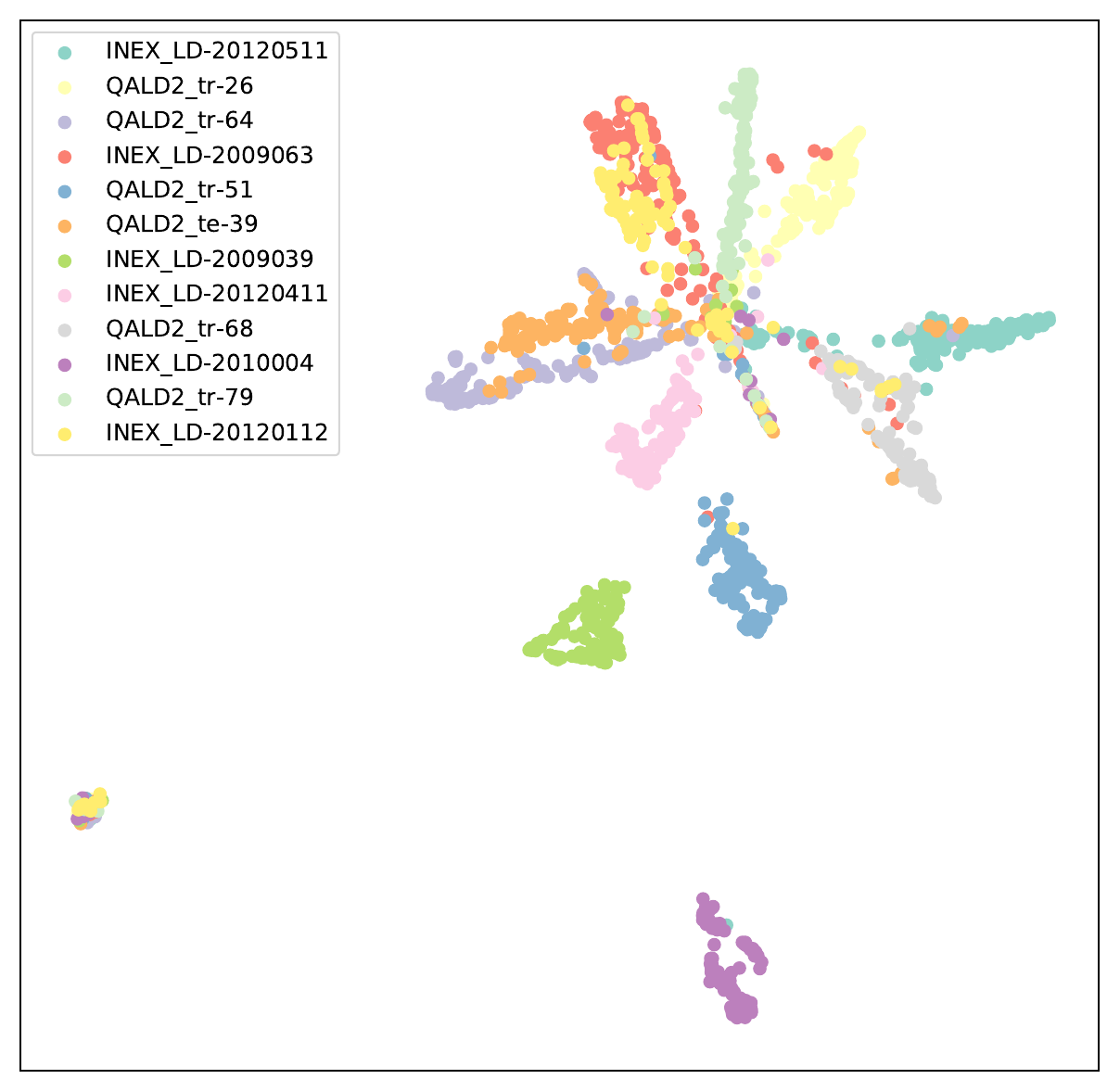}
		\caption{Embeddings without link graph}
		\label{fig:umap:nolg}
	\end{subfigure}
	% \begin{subfigure}[t]{0.48\textwidth}
	% \includegraphics[width=0.98\linewidth]{Figures/umap}
	% 	\caption{With added graph structure}
	% 		\label{fig:umaprel12basicfp}
 %            \end{subfigure}
	% \begin{subfigure}[t]{0.48\textwidth}
 %        \includegraphics[width=\linewidth]{Figures/umap_no_lg}
	% 	\caption{Without added graph structure}
 %        \label{fig:umaprel12nolgfp}
	% \end{subfigure}
	% \label{fig:fig}
	\caption{UMAP visualization of entity embeddings for a subset of queries. Color codes correspond to the relevant entities per query. Queries per code are listed in Table~\ref{tbl:queries} of the Appendix. Default settings of UMAP in python were used.}
	\label{fig:umap}
\end{figure*}

Figure \ref{fig:umap} helps to visually understand how clusters of entities differ for the two embedding variations.
The data points correspond to the entities with a relevance grade higher than 0, for 12 queries
with 100--200 relevant entities in the ground truth data.
We use Uniform Manifold Approximation and Projection (UMAP)~\citep{mcinnes2018umap-software} to reduce the dimensions of the embeddings from 100 to two and plot the projected entities for each query.
In Figure~\ref{fig:umap:nolg}, most of the clusters are overlapping in a star-like shape, while in Figure~\ref{fig:umap:lg}, the clusters are more separated, and the ones with similar search intents are close to each other; e.g., queries \texttt{QALD2\_te-39} and \texttt{QALD2\_tr-64} (which are both about companies), or \texttt{INEX\_LD-20120112} and \texttt{INEX\_LD-2009063} (which are both about war) are situated next to each other.
These analyses support the answer that we formulate for our third research question \textbf{RQ 3:} The graph structure, combined with the textual representation of entities incorporated in Wikipedia2Vec graph embeddings, plays an important role in improving the cluster quality of the representation of the entities and explains the enhanced effectiveness of retrieval results.  
% To observe how false positive entities are placed in the embedding space, we added the 10 highest ranked false positives to the data and created new UMAP plots. \todo{Can we have those plots? If not, remove text?}
% In the obtained plots, false positive entities that are semantically similar to the true positive entities are close to each other. For example, two false positive entities for the query \emph{``South Korean girl groups''} are: \textsc{SHINee} (a South Korean boy band) and \textsc{Hyuna} (a South Korean female singer). Both of these entities are semantically similar to the relevant entities of the query and are also placed in the vicinity of them, although they do not address the information needs of the query. This is consistent with the plots of Figure~\ref{fig:umap} and in line with our conclusion on the effect of graph embeddings for entity-oriented search.

\subsection{Query Analysis}
\label{sec:results:query}

\begin{table}[t]
    \centering
	\caption{Top queries with the highest gains and losses in NDCG at cut-offs 10 and 100, BM25F-CA + Wikipedia2Vec vs.\ BM25F-CA.}
	\begin{tabular}{l || c c}
		\toprule
          \textbf{Query } & \multicolumn{2}{c}{\textbf{Gain in NDCG}} \\
         & @10 & @100 \\
		\hline
		st paul saints                                                    & 0.716 & 0.482	  \\ 
		continents in the world                                           & 0.319 & 0.362	  \\
		What did Bruce Carver die from?                                   & 0.307 & 0.307	   \\
		\hline
		spring shoes canada                                               & -0.286 & -0.286 \\
		vietnam war movie                                                 & -0.470 & -0.240 \\
		mr rourke fantasy island                                          & -0.300 & -0.307 \\
		\bottomrule
	\end{tabular}
	\label{tab:bm25}
\end{table}

\begin{table}[t]
        \centering
	\caption{Top queries with the highest gains and losses in NDCG at
		cut-offs 10 and 100, BM25F-CA + Wikipedia2Vec vs.\ BM25F-CA + Wikipedia2Vec (no graph).}
	\begin{tabular}{l || c c}
		\toprule
          \textbf{Query } & \multicolumn{2}{c}{\textbf{Gain in NDCG}} \\
         & @10 & @100 \\
		\hline
		What did Bruce Carver die from?                                   & 0.307 & 0.307	   \\
		Which other weapons did the designer of the Uzi develop?          & 0.236 & 0.248	  \\ 
		Which instruments did John Lennon play?                           & 0.154 & 0.200	  \\
		
		\hline
		Companies that John Hennessey serves on the board of              & -0.173 & -0.173 \\
		Which European countries have a constitutional monarchy?          & -0.101 & -0.197 \\
		vietnam war movie                                                 & -0.276 & -0.222 \\
		
		\hline
	\end{tabular}
	\label{tab:bm25c}
\end{table}

Next, we analyze queries that are helped and hurt the most by our embedding-based method.
Table~\ref{tab:bm25} shows six queries that are affected the most by
BM25F-CA+ Wikipedia2Vec compared to BM25F-CA, with respect to NDCG@100.
Each of the three queries with the highest gains is linked to at least one
relevant entity (according to the assessments).
The losses can be attributed to various sources of errors.
%For the three entities with the greatest loss, the entity linker is partly at fault. 
%The linked entities to these queries were not all in agreement with the search intent. 
%For the query \emph{spring shoe canada}, there is only relevant entity: the shoe brand \emph{Sully Wong}. This entity is, however, in the 2.4\% of entities without an embedding.
For the query \emph{``spring shoe canada''}, the only relevant entity belongs to the 2.4\% of entities that have no embedding (cf.~\S\ref{sec:exp:training}). 
Query \emph{``vietnam war movie''} is linked to entities
\textsc{Vietnam War} and \textsc{War film}, with confidence scores of
0.7 and 0.2, respectively. This emphasizes Vietnam war facts instead
of its movies, and could be resolved by improving the
accuracy of the entity linker and/or employing a re-ranking approach
that is more robust to linking errors.
The query \emph{``mr rourke fantasy island''} is linked to a wrong
entity due to a spelling mistake\newreview{, which emphasizes the importance of the quality of the entity linker.}
%The entities with the most gain all get linked to at least one relevant entity, according to the relevance assessment.

To further understand the difference between the two versions of the embeddings at the query-level, we selected the queries with the highest and lowest gain in NDCG@100 (i.e., comparing BM25F-CA + Wikipedia2Vec and  BM25F-CA + Wikipedia2Vec (no graph)).
% the scores with embeddings with and without the graph structure. For this again, we look at the most/least helped queries in table~\ref{tab:bm25c}. 
For the query \emph{``Which instruments did John Lennon play?''}, the two linked entities (with the highest confidence score) are \textsc{John Lennon} and \textsc{Musical Instruments}. 
Their closest entity in graph embedding space is \textsc{John Lennon's musical instruments}, relevant to the query. 
This entity, however, is not among the most similar entities when we consider the context-only case. For the other queries in Table \ref{tab:bm25c}, the effect is similar but less prominent than in the BM25F-CA and BM25F-CA + Wikipedia2Vec case, probably due to the lower value of $\lambda$.

%For the other queries, we observe similar behaviour \todo{ ...}

%\todo{The closest relevant entity in graph embedding space is 
%\textsc{John Lennon's musical instruments}. For the context based
%method, this entity is not found, and the closest entities are
%either related to \textsc{John Lennon} or \textsc{Musical Instruments};
%not to both. 

%For \emph{Companies that John Hennessey serves on the board   of}, it links to John F. Hennessey, a tennis player, instead of John L. Hennessy, the computer scientist the query is looking for. 
%For the negative ones, \emph{vietnam war movies} is still in the top 3. This could be because of the lower $\lambda$ value for BM25F + Context, giving less emphasis to the irrelevant entities found by the embedding based method.  
%For the remaining two queries, we could not find a clear cause for why they perform better or worse. But again, the difference could be the value for $\lambda$

\section{Conclusion}
\label{sec:concl}

In this paper, we investigated the use of different types of entity embeddings and different types of entity linkers for entity retrieval. We trained entity embeddings using Wikipedia2Vec, ComplEx, and RDF2Vec, combined these with state-of-the-art entity ranking models, and found empirically that using graph embeddings leads to increased effectiveness of entity retrieval.

We analyzed the effect of different entity linkers and concluded that the most suitable entity linkers are SMAPH, TagMe, and ELQ, annotating both named entities and concepts. When evaluating with the baselines of annotated entities with two different philosophies, multiple scenarios compared to named entities and concepts, we found that SMAPH, TagMe, and ELQ align the most with the annotations focused on named entities and concepts, thus confirming our conclusion that annotated concept are important for retrieval.

We then compared three classes of graph embedding methods, Wikipedia2Vec, RDF2Vec, and ComplEx, and found that first, having as many different entities in the graph embedding will give the best performance, even if they might be redundant. Second, Wikipedia2Vec performs best in all categories, provided that effort is put into solving as many entities linked to an entity without embedding as possible.
Wikipedia2Vec has the highest cluster similarity score, confirming that Wikipedia2Vec is a highly suitable method for performing entity retrieval. 

We conclude that enriching entity retrieval methods with entity embeddings is valuable, efficient, and effective. The choice of entity linker, graph embedding method, and effort to find missing entities are integral to the method's performance. 
\new{For future work, we would like to evaluate how these different graph embedding methods influence more modern Transformer-based entity retrieval methods, as well as how well these methods can be adapted to work on domain-specific entities or sparser knowledge graphs.}

\vskip 0.2in
\bibliography{literature}

\appendix
\newpage
\section{Extra Results}
\shrink
\begin{table}[h]
    \shrink
    \centering
    \small
    \caption{Breakdown per query type of entity retrieval results using different entity linkers and gold annotations. Wikipedia2Vec 2019 is used for re-ranking of BM25F-CA results. Superscripts denote statistically significant differences corresponding to the beginning letter of entity linkers' names, BM25F-CA, and Webis.}%Superscripts denote statistically significant differences (better or worse) corresponding to the beginning letter of the linker.}
    \begin{tabular}{l @{~}|| @{~}l@{~~}l | l@{~~}l | l@{~~}l | l@{~~}l}% | l@{~~}l}
		 \toprule

		 & \multicolumn{2}{c|}{\textbf{SemSearch}} & \multicolumn{2}{c|}{\textbf{INEX-LD}} & \multicolumn{2}{c|}{\textbf{ListSearch}} & \multicolumn{2}{c}{\textbf{QALD-2}}\\% & \multicolumn{2}{c}{\bfseries Total}\\
		%    \hline
		 & @10 & @100 & @10 & @100 & @10 & @100 & @10 & @100 \\%& @10 & @100 \\
		 \hline
 %   \begin{tabular}{l || @{~}l @{~}l |@{~}l @{~}l |@{~}l @{~}l| @{~}l @{~}l |@{~}l @{~}l}
 %    \hline
 %        & \multicolumn{2}{c|}{\textbf{SemSearch}} & \multicolumn{2}{c|}{\textbf{INEX-LD}} & \multicolumn{2}{c|}{\textbf{ListSearch}} & \multicolumn{2}{c|}{\textbf{QALD-2}} & \multicolumn{2}{c}{\bfseries Total}\\
	% 	%    \hline
	% & @10 & @100 & @10 & @100 & @10 & @100 & @10 & @100 & @10 & @100 \\
 %       \hline
    TagMe & 0.661$^{\tiny b}$ & 0.738$^{\tiny b}$ & 0.464$^{\tiny b}$ & 0.552$^{\tiny b}$ & 0.447$^{\tiny b}$ & 0.532$^{\tiny b}$ & 0.387$^{\tiny b}$ & 0.479$^{\tiny b}$ \\%& 0.484$^{\tiny b}$ & 0.570$^{\tiny b}$ \\
    SMAPH & 0.661$^{\tiny b}$ & 0.734 & 0.455 & 0.546$^{\tiny b}$ & 0.448$^{\tiny b}$ & 0.534$^{\tiny b}$ & 0.389$^{\tiny b}$ & 0.483$^{\tiny b}$ \\%& 0.483$^{\tiny b}$ & 0.57$^{\tiny b}$ \\
    Nordlys & 0.645 & 0.719$^{\tiny t}$ & 0.451$^{\tiny b}$ & 0.537$^{\tiny bt}$ & 0.444$^{\tiny b}$ & 0.529$^{\tiny b}$ & 0.387$^{\tiny b}$ & 0.481$^{\tiny b}$ \\%& 0.477$^{\tiny b}$ & 0.563$^{\tiny bts}$ \\
    REL & 0.647$^{\tiny b}$ & 0.732$^{\tiny b}$ & 0.451 & 0.536$^{\tiny t}$ & 0.436 & 0.523$^{\tiny bs}$ & 0.376$^{\tiny bs}$ & 0.472$^{\tiny bsn}$ \\%& 0.472$^{\tiny bts}$ & 0.561$^{\tiny bts}$ \\
    ELQ & 0.645 & 0.729 & 0.474$^{\tiny bnr}$ & 0.554$^{\tiny bnr}$ & 0.455$^{\tiny b}$ & 0.537$^{\tiny b}$ & 0.402$^{\tiny btr}$ & 0.488$^{\tiny br}$ \\%& 0.489$^{\tiny bnr}$ & 0.573$^{\tiny bnr}$ \\
    \hline
    Combined & 0.667$^{\tiny be}$ & 0.738$^{\tiny n}$ & 0.476$^{\tiny btsnr}$ & 0.56$^{\tiny btsnr}$ & 0.468$^{\tiny btsnr}$ & 0.549$^{\tiny btsnr}$ & 0.402$^{\tiny btsnr}$ & 0.492$^{\tiny btsnr}$ \\%& 0.498$^{\tiny btsnre}$ & 0.58$^{\tiny btsnre}$ \\
    \hline
    Webis & 0.63$^{\tiny tsa}$ & 0.711$^{\tiny tsra}$ & 0.452$^{\tiny ea}$ & 0.537$^{\tiny tea}$ & 0.439$^{\tiny a}$ & 0.532$^{\tiny bra}$ & 0.396$^{\tiny bnr}$ & 0.488$^{\tiny br}$ \\%& 0.475$^{\tiny bea}$ & 0.563$^{\tiny btsea}$ \\
    Radboud & 0.651 & 0.727 & 0.478$^{\tiny btsnrw}$ & 0.56$^{\tiny bsnrw}$ & 0.462$^{\tiny bsnrw}$ & 0.543$^{\tiny bsnrw}$ & 0.399$^{\tiny br}$ & 0.495$^{\tiny btsnr}$ \\%& 0.492$^{\tiny bsnrw}$ & 0.577$^{\tiny bnrw}$ \\
    \bottomrule
    \end{tabular}
   \label{tbl:res_el_full}
\shrink
\end{table}

% \begin{landscape}
% \thispagestyle{empty} %removes the page number
\begin{table}[h]
\shrink
\centering
 \small
	 \caption{Breakdown per query type of entity retrieval results using different graph embeddings. Superscripts denote statistically significant differences (better or worse) corresponding to that line of the table.} 
	\label{tbl:res2}
	\begin{tabular}{l @{~}|| @{~}l@{~~}l | l@{~~}l | l@{~~}l | l@{~~}l@{~} }%| l@{~~}l}
		 \toprule

        & \multicolumn{2}{c|}{\textbf{SemSearch}} & \multicolumn{2}{c|}{\textbf{INEX-LD}} & \multicolumn{2}{c|}{\textbf{ListSearch}} & \multicolumn{2}{c}{\textbf{QALD-2}} \\%& \multicolumn{2}{c}{\bfseries Total}\\
		%    \hline
	& @10 & @100 & @10 & @100 & @10 & @100 & @10 & @100 \\%& @10 & @100 \\
		 \hline

	\multicolumn{9}{l}{\emph{Base}}\\
		\hline
		% Wikipedia2Vec (No Graph)     & 0.381 & 0.424 & 0.194 & 0.253 &\textbf{ 0.211} & 0.283 & 0.192 & 0.252 \\%& 0.243 & 0.301 \\ 
	   Wikipedia2Vec 2019 &  \textbf{0.417} & \textbf{0.478} & \textbf{0.217} & \textbf{0.286} &\textbf{ 0.211} & \textbf{0.302} & \textbf{0.212}& \textbf{0.282} \\%& \textbf{0.262}$^{1}$ & \textbf{0.335}$^{1}$\\
		Wikipedia2Vec 2015 & 0.249$^{1}$ & 0.345$^{1}$ & 0.152$^{1}$ & 0.24$^{1}$ & 0.153$^{1}$ & 0.26$^{1}$ & 0.181$^{1}$ & 0.265 \\%& 0.184$^{1}$ & 0.278$^{1}$ \\
		
		ComplEx & 0.309$^{12}$ & 0.322$^{1}$ & 0.136$^{1}$ & 0.17$^{12}$ & 0.139$^{1}$ & 0.178$^{12}$ & 0.148$^{12}$ & 0.193$^{12}$ \\%& 0.182$^{1}$ & 0.216$^{12}$  \\
		ComplEx Pagelinks & 0.32$^{12}$ & 0.371$^{13}$ & 0.148$^{1}$ & 0.187$^{12}$ & 0.183$^{13}$ & 0.239$^{13}$ & 0.168$^{1}$ & 0.224$^{123}$ \\%& 0.204$^{124}$ & 0.255$^{124}$
		RDF2Vec & 0.317$^{12}$ & 0.346$^{1}$ & 0.159$^{1}$ & 0.182$^{12}$ & 0.128$^{14}$ & 0.175$^{124}$ & 0.154$^{1}$ & 0.216$^{12}$ \\%& 0.188$^{15}$ & 0.23$^{1245}$\\
		RDF2Vec Pagelinks & 0.308$^{12}$ & 0.376$^{135}$ & 0.179$^{13}$ & 0.239$^{1345}$ & 0.142$^{14}$ & 0.216$^{1235}$ & 0.158$^{1}$ & 0.225$^{123}$ \\%& 0.195$^{1}$ & 0.262$^{1246}$ \\
		\hline
		\multicolumn{9}{l}{\emph{TagMe}}\\
		\hline

		BM25F-CA         & 0.628 & 0.720 & 0.439 & 0.530 & 0.425 & 0.511 & 0.369 & 0.461 \\%& 0.461 & 0.551 \\
            + Wikipedia2Vec 2019 &  \textbf{0.661}$^{1}$ & \textbf{0.738}$^{1}$ & \textbf{0.464}$^{1}$ &\textbf{ 0.552}$^{1}$ & \textbf{0.447}$^{1}$ & \textbf{0.532}$^{1}$ & \textbf{0.387}$^{1}$ & \textbf{0.479}$^{1}$\\% & \textbf{0.484}$^{1}$ & \textbf{0.57}$^{1}$\\
            + Wikipedia2Vec 2015  & 0.633$^{2}$ & 0.724$^{2}$ & 0.443$^{2}$ & 0.539$^{12}$ & 0.434$^{2}$ & 0.525$^{12}$ & 0.373$^{2}$ & 0.467$^{2}$ \\%& 0.466$^{2}$ & 0.559$^{12}$\\
            + ComplEx & 0.64 & 0.727 & 0.441$^{2}$ & 0.53$^{2}$ & 0.433$^{2}$ & 0.523$^{12}$ & 0.376$^{2}$ & 0.47$^{1}$\\% & 0.468$^{12}$ & 0.558$^{12}$\\
            + ComplEx Pagelinks & 0.653$^{13}$ & 0.732$^{1}$ & 0.445$^{2}$ & 0.537$^{2}$ & 0.443$^{1}$ & 0.532$^{14}$ & 0.377 & 0.474$^{1}$ \\%& 0.474$^{1234}$ & 0.564$^{1234}$ \\
            + RDF2Vec &  0.633$^{25}$ & 0.722$^{25}$ & 0.429$^{2345}$ & 0.53$^{23}$ & 0.428$^{25}$ & 0.519$^{125}$ & 0.374$^{2}$ & 0.465$^{2}$ \\%& 0.461$^{245}$ & 0.554$^{235}$ \\
            + RDF2Vec Pagelinks & 0.636$^{25}$ & 0.727 & 0.436$^{2}$ & 0.539$^{126}$ & 0.432$^{2}$ & 0.52$^{125}$ & 0.381$^{1}$ & 0.474$^{136}$ \\%& 0.467$^{256}$ & 0.56$^{126}$ \\
		\hline
		\multicolumn{9}{l}{\emph{Concepts}}\\
		\hline
   		BM25F-CA         & 0.628 & 0.720 & 0.439 & 0.530 & 0.425 & 0.511 & 0.369 & 0.461 \\%& 0.461 & 0.551 \\
            + Wikipedia2Vec 2019 & 0.652 & 0.73 & 0.48$^{1}$ & 0.558$^{1}$ & 0.46$^{1}$ & 0.543$^{1}$ & 0.398$^{1}$ & 0.491$^{1}$ \\%& 0.492$^{1}$ & 0.576$^{1}$ \\
            + Wikipedia2Vec 2015 & 0.632$^{2}$ & 0.721 & 0.449$^{2}$ & 0.541$^{12}$ & 0.442$^{12}$ & 0.531$^{12}$ & 0.387$^{12}$ & 0.479$^{12}$ \\%& 0.473$^{12}$ & 0.564$^{12}$ \\
            + ComplEx & 0.635 & 0.723 & 0.444$^{2}$ & 0.535$^{2}$ & 0.434$^{2}$ & 0.522$^{23}$ & 0.375$^{23}$ & 0.47$^{123}$ \\%& 0.467$^{12}$ & 0.558$^{123}$ \\
            + ComplEx Pagelinks & 0.65$^{14}$ & 0.727 & 0.449$^{2}$ & 0.539$^{12}$ & 0.44$^{2}$ & 0.53$^{124}$ & 0.384$^{1}$ & 0.477$^{12}$ \\%& 0.476$^{124}$ & 0.564$^{124}$ \\
            + RDF2Vec & 0.637$^{5}$ & 0.723 & 0.429$^{2345}$ & 0.529$^{235}$ & 0.43$^{23}$ & 0.52$^{235}$ & 0.373$^{235}$ & 0.47$^{1235}$ \\%& 0.463$^{235}$ & 0.556$^{1235}$ \\
            + RDF2Vec Pagelinks & 0.637 & 0.733$^{346}$ & 0.443$^{26}$ & 0.543$^{1246}$ & 0.437$^{2}$ & 0.526$^{12}$ & 0.384$^{126}$ & 0.482$^{146}$ \\%& 0.471$^{126}$ & 0.566$^{1246}$\\
		\bottomrule
	\end{tabular}
        \label{tbl:res_emb_full}
\end{table}

\newpage
\section{Queries}\label{app:queries}

%Table \ref{tbl:queries} lists the queries referred to by their
%DBpedia-Entity V2 identifier (in Figures and/or text).
\shrink
\begin{table}[h]
  \caption{Queries used for Figure~\ref{fig:umap}.}\label{tbl:queries}
  % \scriptsize
  \begin{tabular}{l||l}
    \toprule
    \textbf{Query ID} & \textbf{Query text} \\
    \midrule
    INEX\_LD-20120511 & female rock singers \\
    QALD2\_tr-26 & Which bridges are of the same type as the Manhattan Bridge? \\
    QALD2\_tr-64 & Which software has been developed by organizations founded in California? \\
    INEX\_LD-2009063 & D-Day normandy invasion \\
    QALD2\_tr-51 & Give me all school types. \\
    QALD2\_te-39 & Give me all companies in Munich. \\
    INEX\_LD-2009039 & roman architecture \\
    INEX\_LD-20120411 & bicycle sport races \\
    QALD2\_tr-68 & Which actors were born in Germany? \\
    INEX\_LD-2010004 & Indian food \\
    QALD2\_tr-79 & Which airports are located in California, USA? \\
    INEX\_LD-20120112 & vietnam war facts \\
    \bottomrule
  \end{tabular}
  
\end{table}

\end{document}